\documentclass[conference]{IEEEtran}

\usepackage{booktabs} 
\usepackage[utf8]{inputenc}
\usepackage[titlenumbered,ruled]{algorithm2e}
\usepackage{color}
\usepackage{subfigure}
\usepackage{comment}
\usepackage{array}
\usepackage{lineno,hyperref}
\usepackage[justification=centering]{caption}
\usepackage{listings}
\usepackage{float}
\usepackage{graphicx}
\usepackage{booktabs}
\usepackage{longtable}
\usepackage{multirow}
\usepackage{lipsum}
\usepackage{url}
\usepackage{verbatim}
\usepackage[flushleft]{threeparttable}

\usepackage{amsmath,amssymb,amsfonts}
\usepackage{bm}
\usepackage{mathtools}
\usepackage{calrsfs}
\usepackage[mathscr]{euscript}

\title{Resilient and Adaptive Framework for Large Scale Android Malware Fingerprinting using Deep Learning and NLP Techniques}

\author{
\IEEEauthorblockN{{\bf ElMouatez Billah Karbab}}
\IEEEauthorblockA{{\it Concordia University}\\{\tt elmouatez.karbab@concordia.ca}}
\and
\IEEEauthorblockN{{\bf Mourad Debbabi}}
\IEEEauthorblockA{{\it Concordia University}\\ {\tt mourad.debbabi@concordia.ca}}
}

\begin{document}
\maketitle

\begin{abstract}

  Android malware detection is a significant problem that affects billions of
users using millions of Android applications (apps) in existing markets. This
paper proposes \textsf{PetaDroid}, a framework for accurate Android malware
detection and family clustering on top of static analyses. \textsf{PetaDroid}
automatically adapts to Android malware and benign changes over time with
resilience to common binary obfuscation techniques. The framework employs novel
techniques elaborated on top of natural language processing (NLP) and machine
learning techniques to achieve accurate, adaptive, and resilient Android malware
detection and family clustering. \textsf{PetaDroid} identifies malware using an
ensemble of convolutional neural network (CNN) on proposed Inst2Vec features.
The framework clusters the detected malware samples into malware family groups
utilizing sample feature digests generated using deep neural auto-encoder. For
change adaptation, \textsf{PetaDroid} leverages the detection confidence
probability during deployment to automatically collect extension datasets and
periodically use them to build new malware detection models. Besides,
\textsf{PetaDroid} uses code-fragment randomization during the training to
enhance the resiliency to common obfuscation techniques. We extensively
evaluated \textsf{PetaDroid} on multiple reference datasets.
{\sf PetaDroid} achieved a high detection rate (98-99\%
f1-score) under different evaluation settings with high homogeneity in the
produced clusters (96\%). We conducted a thorough quantitative comparison with
state-of-the-art solutions {\sf MaMaDroid}, {\sf DroidAPIMiner}, {\sf MalDozer},
in which {\sf PetaDroid} outperforms them under all the evaluation settings.

\end{abstract}

\section{Introduction}

Android OS's popularity has increased tremendously since the last decade. It is
undoubtedly an appropriate choice for smart mobile devices such as phones and
tablets or the internet of things devices such as TVs due to its open-source
license and the massive number of useful apps developed for this platform
(about 4 Million apps in 2019 \cite{Statcounter}). Nevertheless, malicious apps
target billions of Android users through centralized app markets. The detected
malicious apps increased by 40\% in 2018-Q3 compared to the same period in 2017
\cite{GDATA}. Google Play \cite{oberheide2012dissecting} employs a vetting
system named \texttt{Bouncer} to detect malicious apps through static and
dynamic analyses. Despite these analyses, many malicious
apps\footnote{https://tinyurl.com/y4qdtuy9} were able to bypass
\texttt{Bouncer} and infect several hundred thousand
devices\footnote{https://tinyurl.com/y4mckwxm}. Therefore, there is a dire need
for accurate, adaptive, yet resilient Android malware detection systems for the
app market scale.

\subsection{Problem Statement}

In this paper, we identify the following gaps in the state-of-the-art solutions
for Android malware detection:

{\bf P1:} The accuracy of Android malware detection systems tends to decrease
over time due to different factors: (1) variations in existing malware family,
(2) new malware families, (3) and new Android APIs in benign and malicious apps.
These factors are mostly reflected in the changes in Android API call sequences
in malicious and benign apps. Nevertheless, these changes are incremental in
most cases compared to the existing apps. In this context, we consider two
problems: (1) The resiliency of the detection systems that use machine learning
models \cite{mariconti2019mamadroid} to changes over time, (2) and the
possibility of automatic adaptation to the new changes
\cite{DBLP:conf/eurosp/XuLDCX19}.


{\bf P2:} Android malware family attribution is an important problem in the
realm of malware detection. The malware family attribution could be important
essential to define the threats\footnote{https://tinyurl.com/yydg5vew} of the
detected malware \cite{DBLP:conf/malware/MassarelliACQUB17}. However, few
existing solutions
\cite{DBLP:journals/cn/BaiXMLF21,DBLP:journals/compsec/ZhangTC19} provide
Android malware family attribution. Furthermore, these solutions rely on
supervised learning where prior knowledge of the families is required
\cite{DBLP:journals/tosem/GarciaHM18}. However, such knowledge is hard to get
and not realistic in many cases, especially for new malware
families\footnote{https://tinyurl.com/y8rc6q89}.

{\bf P3:} malware developers employ various obfuscation techniques to thwart
detection attempts. Obfuscation resiliency is a key requirement in modern
malware fingerprinting that applies static analyses. Few solutions address
the obfuscation issue in the context of Android malware detection, more
specifically, the resiliency to common obfuscations and binary code
transformations.

{\bf P4:} Many existing solutions \cite{arp2014drebin, suarez2017droidsieve,
Sen2016StormDroid} rely on manual feature engineering to extract relevant
malware patterns for classification. Despite the good detection performance,
manual feature engineering is not scalable to the amount and the growth pace of
Android malware. Therefore, there is a dire need for malware detection
solutions that rely on automatic feature engineering to discover relevant
malware patterns. Other solutions \cite{mariconti2019mamadroid} utilize
automated feature engineering but lack efficiency. The malware detection
process's efficiency is a paramount requirement due to the growing number of
Android apps.

\subsection{Proposed Solution}

In this paper, we propose \textsf{PetaDroid}, an accurate, adaptive, resilient,
and yet efficient Android malware detection and family clustering using natural
language processing (NLP) and deep learning techniques on top of static
analysis features. In \textsf{PetaDroid}, we aim to address the previously
mentioned problems as follows:

{\bf 1.} Our fundamental intuition for time resiliency and adaptation is that
Android apps are changing over time incrementally. Benign apps embrace new
Android APIs, deprecations, and components gracefully to do not disturb the
user experience. Malware developers aim to target the maximum devices by
employing stable and cross-Android version APIs. We argue that
\textsf{PetaDroid} can fingerprint malicious apps within a time window with
high confidence because the application still contains enough patterns of
similarity to known samples. In this period, we argue that those top confidence
detection apps (malicious or benign) could extend our initial build dataset of
\textsf{PetaDroid} to enhance the overall performance and keep up with the
change over time. \textsf{PetaDroid} introduces an automatic and continuous
dataset enrichment technique and machine learning model training to overcome
the change over time. 

{\bf 2.} \textsf{PetaDroid} goes a step further in the detection process by
clustering the detected samples into groups with high similarity. We {\it
exclusively} group highly similar samples, most likely of the same malware
family. \textsf{PetaDroid} family attribution is found upon the assumption that
malicious applications tend to have similar characteristics in the Android
Dalvik bytecode code.  For example, SMS malicious apps exploit the same Android
APIs in a very similar manner. The similarity increases when malicious apps are
from the same malware family. We leverage this assumption to build an automatic
and unsupervised malware family tagging system using deep neural network
auto-encoder for sample digest generation on top of \textit{InstNGram2Bag}
features (based on NLP bag of words). Using the DBScan
\cite{DBLP:conf/kdd/EsterKSX96} clustering algorithm, we cluster the {\it most
similar samples} from the detected malicious apps.

{\bf 3.} {\sf PetaDroid} introduces code fragments randomization during
training and deployment phases to enhance the obfuscation resiliency. We
artificially apply random permutations to change the order of code basic-blocks
without altering the basic-block instructions. We consider a code basic-block
as a possible micro-action in the app execution flows. Therefore, we randomize
the app execution flows without affecting the micro-actions within the flow to
emulate code transformation during the training and deployment phases. Code
fragment randomization strengthens the obfuscation robustness of
\textsf{PetaDroid}, as shown in Section \ref{sec_petaObfusResiliency}.

{\bf 4.} \textsf{PetaDroid} leverages deep learning techniques to achieve
automatic feature engineering. Specifically, We employ an ensemble of CNN
models on top of canonical instruction embeddings, namely Inst2Vec. Those
embeddings are learned using word2vec \cite{Mikolov2013Distributed}, an NLP
technique that translates latent patterns and semantics from raw word sequences
into word embeddings. \textsf{PetaDroid} detection is very efficient due to its
minimal preprocessing compared to state-of-the-art solutions
\cite{mariconti2019mamadroid}.


\subsection{Contributions and Outline}
The main contributions of this paper are:

(1) We propose a novel adaptation technique for Android malware detection to
automatically adapt the detection system. The proposed techniques rely on the
confidence probability of the detection ensemble to collect extension training
datasets from received samples during the deployment (Section
\ref{sec_appraochExtensionUpdate}). 

(2) We propose a novel fragment randomization technique to boost the detection
system resiliency to common code-obfuscation techniques. In this technique, we
randomize the order of code basic-blocks without affecting the basic-blocks
instructions during the training and the deployment phases (Section
\ref{sec_appraochFragementDetection}). 

(3) We propose \textsf{PetaDroid}, an accurate and efficient malware detection
and clustering framework based on code static analyses, NLP, and machine
learning techniques. In \textsf{PetaDroid}, we propose an ensemble of CNN
models on top of a code embedding model, namely {\it Inst2Vec}, to accurately
detect malware with probability confidence
(Section~\ref{sec_appraochMalwareDetection}). Besides, we propose an
unsupervised family attribution by clustering malware family samples on top of
{\it InstNgram2Bag} features generated using deep auto-encoders
(Section~\ref{sec_appraochClustering}). We released the source code of {\sf
PetaDroid} to the community in \url{https://github.com/mouatez/petadroid}.

(4) We extensively evaluate \textsf{PetaDroid} to assess its effectiveness and
efficiency on different reference datasets of \textsf{PetaDroid} under various
evaluation settings (Section \ref{sec_appraochAppRepresentation}). To
demonstrate the framework robustness against common obfuscation techniques, we
evaluate \textsf{PetaDroid} on the obfuscated dataset generated using
DroidChameleon \cite{Vaibhav2013DroidChameleon} obfuscation tool and PRAGuard
\cite{Davide2015Praguard} dataset (Section \ref{sec_petaObfusResiliency}). We
conduct an empirical comparison study between state-of-the-art solutions, namely
\textsf{MaMaDroid} \cite{mariconti2019mamadroid}, \textsf{DroidAPIMiner}
\cite{Aafer2013DroidAPIMiner}, and {\sf MalDozer} \cite{karbab2018maldozer} in
which \textsf{PetaDroid} outperforms these solutions (Section
\ref{sec_compStudy}).

\section{PetaDroid}
\begin{figure}
\centering
    \includegraphics[width=0.50\textwidth, trim=0.1cm 7.9cm 6.1cm 0.2cm, clip]{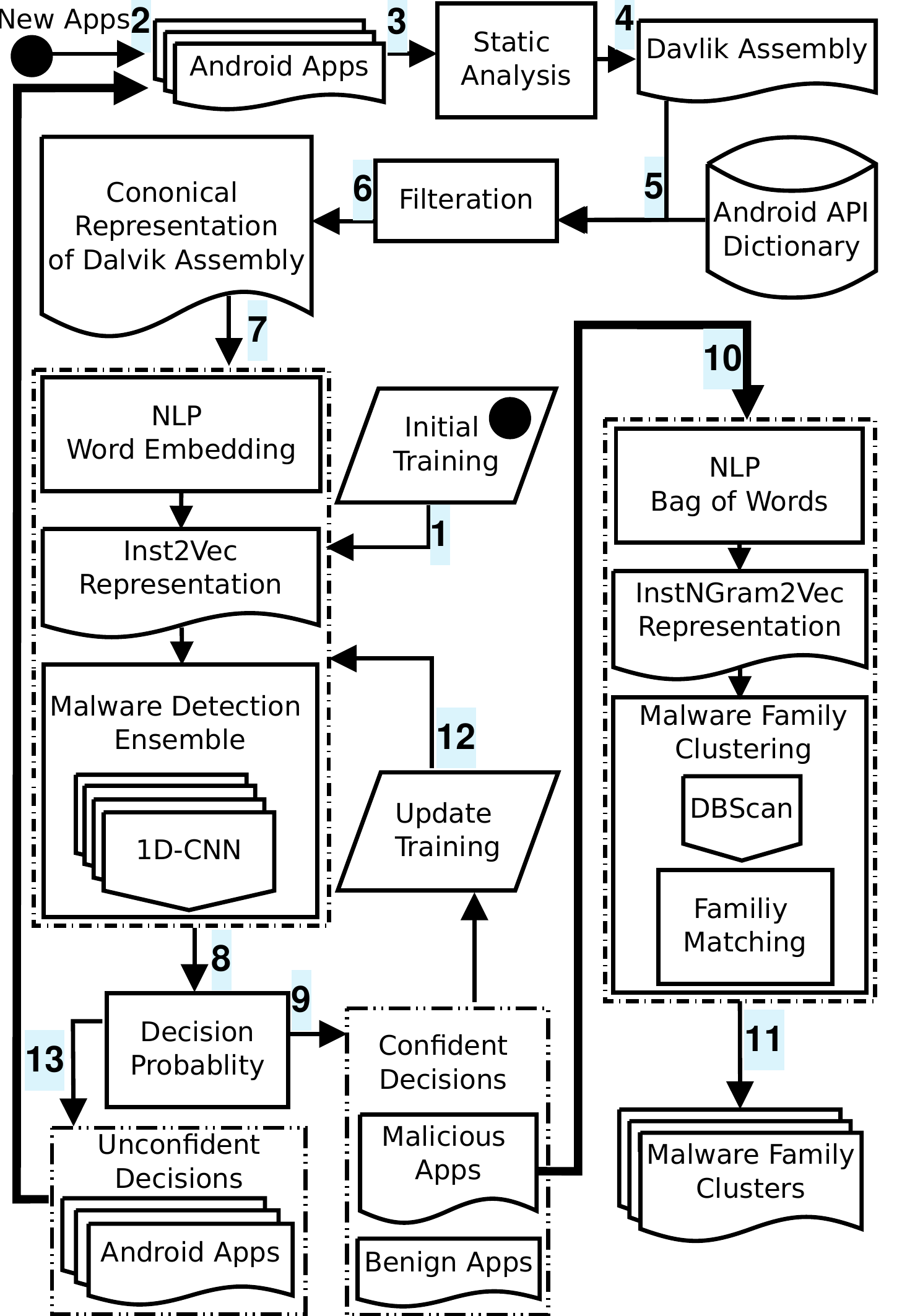}
\caption{PetaDroid Architecture Overview}
\label{fig_petadroid_approach}
\end{figure}

In this section, we detail \textsf{PetaDroid} methodology and its components.
The general overview of \textsf{PetaDroid} is presented in Figure
\ref{fig_petadroid_approach}.

-------------------------------
\subsection{Methodology Summary} 

\textsf{PetaDroid} employs static analyses on Android Packaging (APK) to
investigate the maliciousness of Android apps. \textsf{PetaDroid} starts by
extracting raw static features from the Android Packaging, specifically the
Dalvik bytecode. We develop a fast preprocessing phase to extract raw Dalvik
assembly instructions. We generate, on the fly, the canonical form of the
assembly instructions by substituting the value of constants, memory addresses
with symbolic names. The output is a raw sequence of canonical assembly
instructions of Dalvik virtual machine.

\textsf{PetaDroid} maintains a logical separation among the software component
of an Android app. We keep track of methods' instruction sequences within the
app global instruction sequence. It is a natural breakdown because an Android
app is a set of classes, and a class is a set of methods and attributes. {\it In
the context of Android apps, we consider the class's methods as our basic-block
codes, which we keep unaltered when applying the fragment randomization technique.
Our tests for granularity beyond methods (.i.e code basic-blocks of the method'
Control Flow Graph (CFG)) results in very small code basic-blocks (one
instruction in many cases)}. The global app execution sequence is composed of a
list of micro-execution paths (instruction sequences of the methods), through
which the execution proceeds during runtime. The extracted canonical instruction
sequences (described in the next section) help preserve the underlying
micro-execution paths while not emphasizing the global execution order.

Previous solutions \cite{mariconti2019mamadroid} apply heavy and complex
preprocessing to construct a global call graph to simulate runtime execution
using static analyses. In contrast, our extraction approach is lightweight
because we consider only the class's methods. This allows swift preprocessing in
commodity hardware while maintaining the intended granularity. Furthermore, and
in contrast with previous solutions \cite{arp2014drebin,
Aafer2013DroidAPIMiner},  we adopt granular features using a canonical
instruction followed by representation learning. We propose custom code modeling
techniques for representation learning inspired by advanced NLP techniques.
Specifically, we design and develop \textit{Inst2Vec} and \textit{InstNGram2Bag}
code modeling techniques to model and discover latent information. In a
nutshell, \textsf{PetaDroid} detection process has the following main
components:

\noindent (1) \textbf{Representation Learning for Classification}:
\textsf{PetaDroid} learns latent representations in an unsupervised manner using
the word2vec technique \cite{Mikolov2013Distributed} on raw Dalvik canonical
instruction sequences and produces embeddings.

\noindent (2) \textbf{Malware Detection}: \textsf{PetaDroid} employs neural
network models for automatic features engineering from the embedding
representation. During the training phase, the training dataset automatically
guides the feature engineering to discover relevant malware patterns.
\textsf{PetaDroid} detection system rests on the CNN models ensemble that
consumes \textit{Inst2Vec} embedding features and produces maliciousness
probability likelihood.

\noindent (3) \textbf{Model Adaptation}: \textsf{PetaDroid} collects Android
apps that are detected with high confidence, whether malicious or benign, to
extend the {\sf PetaDroid} primary labeled dataset (used to build the current
detection ensemble). Periodically, \textsf{PetaDroid} employs the collected
datasets (primary and extensions) to build new model ensembles to adapt to new
benign and malware patterns overtime automatically.

\noindent (4) \textbf{Code Representation for Clustering}: In contrast with
classification, the malware family clustering requires another representation
that squashes a canonical instruction sequence into one feature vector instead
of a sequence of embeddings for a given malware sample to fit with our
clustering system. For each detected malware sample,  we produce feature vectors
using a bag of words (N-grams) NLP model and the feature hashing technique
\cite{qinfeng09hashk} on top of the canonical instruction sequences. We call the
outcomes InstNGram2Bag vectors, one InstNGram2Bag vector for each detected
malware.

\noindent (5) \textbf{Digest Generation}: we produce malware digests by applying
deep neural auto-encoders \cite{DBLP:conf/icann/HintonKW11} on the InstNGram2Bag
vectors to produce even more compact embedding or a digest for each malware
sample.

\noindent (6) \textbf{Malware Family Clustering}: \textsf{PetaDroid} clusters
the flagged malicious apps into groups with high inter-similarity between their
digests, and most likely of the same malware family. \textsf{PetaDroid}
clustering system is based on DBScan \cite{DBLP:conf/kdd/EsterKSX96} clustering
algorithm.

\subsection{Android App Representation}
\label{sec_appraochAppRepresentation}

In this section, we present the preprocessing of Dalvik code and its
representation into a canonical instruction sequence. We seek the preservation
of the maximum information about apps' behaviors while keeping the process very
efficient. The preprocessing begins with the disassembly of an app bytecode to
Dalvik assembly code, as depicted in Figure \ref{fig_methodSnippet}.

\begin{figure}[h]
\centering
\begin{scriptsize}
\noindent\fbox{%
\parbox{0.45\textwidth}{%
// Object Creation

new-instance v10, \textbf{java/util/HashMap}

// Object Access

invoke-direct {v10}, \textbf{java/util/HashMap}

if-eqz v9, 003e

..

// Method Invocation

// * = \textbf{Android/telephony}

invoke-virtual {v4},
\textbf{*/TelephonyManager.getDeviceId()}\textbf{java/lang/String}

move-result-object v11

// Method Invocation

invoke-virtual {v4},
\textbf{*/TelephonyManager.getSimSerialNumber()}\textbf{java/lang/String}

move-result-object v13

// Method Invocation

invoke-virtual {v4}
\textbf{*/TelephonyManager.getLine1Number()}\textbf{java/lang/String}

move-result-object v4

...

// Object Creation

new-instance v20, \textbf{java/io/FileReader}

const-string v21, "/proc/cpuinfo"

invoke-direct/range {v20, v21},
\textbf{java/io/FileReader.init(java/lang/String)}

new-instance v21, \textbf{java/io/BufferedReader}

...

move/from16 v2, v20

// Field Access

// * = \textbf{Android/content/pm}

iget-object v0, v0,
\textbf{*/ApplicationInfo.metaData}
\textbf{Android/os/Bundle}

move-object/from16 v19, v0
}%
}
\end{scriptsize}

\caption{Android Assembly from a Malware Sample}
\label{fig_methodSnippet}
\end{figure}

We model the Dalvik assembly code as code fragments where each fragment is a
class's method code in the Dalvik assembly. It is a natural separation because
Dalvik code $D$ is composed of a set of classes $D=\{C_1, C_2, \dots C_s\}$.
Each class $C_i$ contains a set of methods $C=\{M_1, M_2, \dots M_k\}$, where
we find actual assembly code instructions. We preserve the order of Dalvik
assembly instructions within methods while ignoring the global execution paths.
Method execution is a possible \textit{micro-behavior} for an Android app,
while a global execution path is a likely \textit{macro-behavior}. An Android
app might have multiple global execution paths based on external events. In
contrast, Android malware tends to have one crucial global execution path
(malicious payload) and other ones to distract malware detection systems. The
malware could produce variations for the payload global execution path.
However, it still depends on the micro-behavior to produce another global one.
\textsf{PetaDroid} assembly preprocessing produces a multiset of sequences
$P=\{S_1, S_2, \dots S_h\}$ where each sequence $S$ contains an ordered
instruction sequence $S=\langle I_1, I_2, \dots I_v \rangle$ of a class's
method. In other words, $P$ contains instruction sequences
$P=\{
\langle I_1, I_2, \dots \rangle_1,
\langle I_1, I_2, \dots \rangle_2,
\dots
\langle I_1, I_2, \dots \rangle_h
\}$
where the order is only preserved inside individual sequences $S_{i}$ (the
methods instructions).  Thus, a sequence $S$ defines a possible micro-execution
(or behavior) from the Android app's overall runtime execution.

\begin{figure}[h]
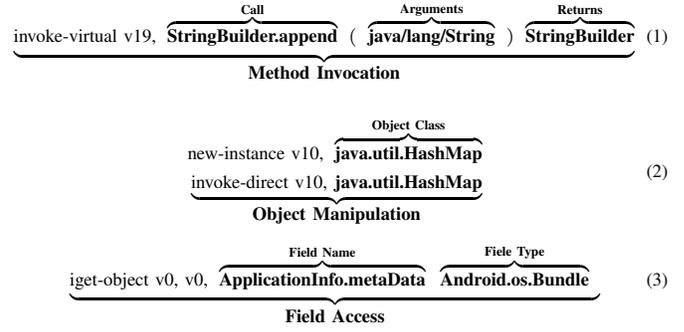

\begin{equation}
\centering 
\scriptsize
\underbrace{
\mbox{invoke-virtual {v19},}
\mbox{\hspace{0.1cm}}
\overbrace{\mbox{\bf StringBuilder.append}}^\text{\bf Call}
\mbox{\hspace{0.1cm}}
(
\mbox{\hspace{0.1cm}}
\overbrace{\mbox{\bf java/lang/String}}
^\text{\bf Arguments}
\mbox{\hspace{0.1cm}}
)
\mbox{\hspace{0.1cm}}
\overbrace{\mbox{\bf StringBuilder}}^\text{\bf Returns}
}_\text{\bf Method Invocation}
\end{equation}

\begin{equation}
\centering 
\scriptsize
\begin{split}
\mbox{new-instance v10,}
\mbox{\hspace{0.1cm}}
\overbrace{\mbox{\bf java.util.HashMap}}^\text{\bf Object Class}
\\\underbrace{
\mbox{invoke-direct {v10},}
\mbox{\hspace{0.1cm}}
\mbox{\bf java.util.HashMap}
}_\text{\bf \scriptsize Object Manipulation}
\end{split}
\end{equation}

\begin{equation}
\centering 
\scriptsize
\underbrace{
\mbox{iget-object v0, v0,}
\mbox{\hspace{0.1cm}}
\overbrace{\mbox{\bf ApplicationInfo.metaData}}^\text{\bf Field Name}
\mbox{\hspace{0.1cm}}
\overbrace{\mbox{\bf Android.os.Bundle}}
^\text{\bf Fiele Type}
\mbox{\hspace{0.1cm}}
}_\text{\bf \scriptsize Field Access}
\end{equation}

\caption{Canonical Representation of Dalvik Assembly}
\label{fig_canonicalRepresentation}
\end{figure}

As shown in Figure \ref{fig_methodSnippet}, the Dalvik assembly is too sparse.
We want to keep the assembly instruction skeleton that reflects possible
runtime behaviors with less sparsity. In \textsf{PetaDroid}, we propose a
canonical representation for Dalvik assembly code, as shown in Figure
\ref{fig_canonicalRepresentation}. The key idea is to keep track of the Android
platform APIs and objects utilized inside the method assembly. To fingerprint
malicious apps, the canonical representation will mostly preserve the actions
and the manipulated system objects, such as sending SMS action or getting
(setting) sensitive information objects. \textsf{PetaDroid} canonical
representation covers three types of Dalvik assembly instructions, namely:
\textit{Method invocation}, \textit{object manipulation}, and \textit{field
access}, as shown in Figure \ref{fig_canonicalRepresentation}. In the method
invocation, we focus on the method call, \textit{Package.ClassName.MethodName},
the parameters list, \textit{Package.ClassName}, and the return type,
\textit{Package.ClassName}. In object manipulation, we capture the class
object, \textit{Package.ClassName}, that is being used. Finally, we track the
access to system fields by capturing the field name,
\textit{Package.ClassName.FieldName}, and its type, \textit{Package.ClassName}.
Our manual inspections of  Dalvik assembly for hundreds of malicious and benign
samples shows that these three forms cover the essential of Dalvik assembly
instructions.

{\sf PetaDroid} instruction parser keeps only the canonical representation and
ignores the rest. For example, our experiments show that Dalvik opcodes add a
lot of sparsity without enhancing the malware fingerprinting performance. On
the contrary, it could negatively affect overall performance, which is shown in
previous solutions \cite{mclaughlin2017deep}. The final step in preprocessing a
method $M$ (see Figure \ref{fig_methodSnippet}) is to flatten the canonical
representation of a method into a single sequence $S$ (see Figure
\ref{fig_flattenCanonical}).  We keep only the Android platform related assets
like API, classes, and system fields in the final method's sequence $S$. We
maintain a vocabulary dictionary (key: value) in the form of $(Android assets:
identifier)$ (for example $( Android/telephony/TelephonyManager: 439 )$ ) of
all Android OS assets (all versions) to filter and map Android assets to unique
identifiers (unique integer for a given Android assets) for the method
instruction sequence during the preprocessing. The output of the app
representation phase is a list of sequences $\hat{P}=\{S_{c1}, cS_{c2}, \dots
cS_{ch}\}$.  Each sequence is an ordered canonical instruction representation of
one method.

\begin{figure}[t]
\centering
\begin{scriptsize}
\noindent\fbox{%

\parbox{0.49\textwidth}{%

\texttt{java/util/HashMap}

\texttt{java/util/HashMap}

..

\texttt{Android/telephony/TelephonyManager.getDeviceId()}

\texttt{java/lang/String}

\texttt{Android/telephony/TelephonyManager.getSimSerialNumber()}

\texttt{java/lang/String}

\texttt{Android/telephony/TelephonyManager.getLine1Number()}

\texttt{java/lang/String}

...

\texttt{java/io/FileReader}

\texttt{java/io/FileReader.init()}

\texttt{java/lang/String}

\texttt{java/io/BufferedReader}

...

\texttt{Android/content/pm/ApplicationInfo.metaData}

\texttt{Android/os/Bundle}

}%
}
\end{scriptsize}

\caption{Flatten Canonical Representation}
\label{fig_flattenCanonical}
\end{figure}

In the following, we summarize the notations used in the rest of the paper:

\begin{table*}[h]
\centering
\caption{Notation Summary}
\label{tab_notationSummary}
\begin{scriptsize}
\centering
\begin{tabular}{| c | p{6cm} | l |}
\hline
{\bf Notation} & {\bf Description} & {\bf Format} \\
\hline
$D$      & Dalvik assembly code of one Android App                            & Raw text                            \\
\hline
$C$      & Dalvik Java Class                                                  & Raw text                            \\
\hline
$M$      & Dalvik Java Method                                                 & Raw text                            \\
\hline
$S$      & Sequence of extracted instructions of one Dalvik Java Method $M$   & List of Dalvik raw text instructions\\
\hline
$P$      & Multiset of methods' sequences $S$                                     & Multiset of sequences                   \\
\hline
$S_{c}$     & Sequence of canonical instructions generated from $S$ using $V$    & List of canonical instruction IDs   \\
\hline
$\hat{P}$ & Multiset of methods' sequences $S_{c}$                                    & Multiset of sequences                   \\
\hline
$P_{c}$     & The result of shuffling and concatenating of all $S_{c}$        & Sequence of canonical instructions      \\
\hline
$F$      & Fragment is a truncated portion from $P_{c}$                          & List of canonical instructions      \\
\hline
$CNNModel$     & Classification model based on Convolutional Neural Network (CNN)   & Deep learning model                 \\
\hline
  $\Phi$    & Ensemble of classification models $\Phi=\{
CNNModel_1, CNNModel_2, \dots CNNModel_{\phi} \}$                             & Set of deep learning models        \\
\hline
$y$      & Dataset label                                                      & Malware or not                      \\
\hline
$\hat{y}$ & Prediction likelihood of the classification models $\hat{y}=\Phi(F)$ & Probability                         \\
\hline
$\zeta$  & Detection threshold for the general decision strategy              & Probability threshold               \\
\hline
$\eta$   & Detection threshold for the confidence decision strategy           & Probability threshold               \\
\hline
\end{tabular}
\end{scriptsize}
\end{table*}

\subsection{Malware Detection}
\label{sec_appraochMalwareDetection}

In this section, we present the {\sf PetaDroid} malware detection process using
CNN on top of \textit{Inst2Vec} embedding features. The detection process starts
from a multiset of discretized canonical instruction sequences
$\hat{P}=\{S_{c1}, S_{c2}, \dots S_{ch}\}$. Notice that $\hat{P}$ is a multiset and not a
set since it might contain duplicated sequences. The duplication comes from
having the same Dalvik method's code in two (or more) distinct Dalvik classes.
\textsf{PetaDroid} CNN ensemble produces a detection result together with
maliciousness and benign detection probabilities for a given sample. To achieve
automatic adaptation, we leverage the detection probabilities to automatically
collect an extension dataset that \textsf{PetaDroid} employs to build new CNN
ensemble models.

\begin{figure}

\begin{equation}
\begin{small}
\hat{P}=\{S_{c1}, S_{c2}, S_{c2}\}
\end{small}
\label{equ_}
\end{equation}

\begin{equation}
\begin{small}
\{
\underbrace{
S_{c3}, S_{c1}, S_{c2}
}_\text{\bf Shuffled sequences}
\}
\end{small}
\label{equ_}
\end{equation}

\begin{equation}
\begin{small}
\{
\overbrace{
\langle I_{1}, I_{1}, \dots I_{|Sc3|}\rangle_3,
\langle I_{1}, I_{1}, \dots I_{|Sc1|}\rangle_1,
\langle I_{1}, I_{1}, \dots I_{|Sc2|}\rangle_2,
}^\text{\bf while preserving the order inside sequences}
\}
\end{small}
\label{equ_}
\end{equation}

\begin{equation}
\begin{small}
\hat{P_{c}}=\{
\overbrace{
\underbrace{
I_{3,1}, I_{3, 1}, \dots I_{3, |Sc3|},
I_{1, 1}, I_{1, 1},
\dots I_{1, |Sc1|},
I_{2, 1},
}_\text{\bf Fragment truncation }
\underbrace{
I_{2, 1}, \dots I_{2, |Sc2|},
}_\text{\bf Rest}
}^\text{\bf Sequence concatenation}
\}
\end{small}
\label{equ_}
\end{equation}

\begin{equation}
\begin{small}
F=\{
\overbrace{
I_{3,1}, I_{3, 1}, \dots I_{3,|Sc3|},
I_{1,1}, I_{1,1},  \dots I_{1,|Sc1|},
I_{2,1}
}^\text{\bf on fragment size $|F|$ }
\}
\end{small}
\label{equ_}
\end{equation}

\caption{Example of Fragment Generation}
\label{fig_exampleFragmentGeneration}
\end{figure}
\subsubsection{Fragment Detection}
\label{sec_appraochFragementDetection}

Fragment-based detection is a key technique in \textsf{PetaDroid}. A fragment
$F$ is a truncated portion from the beginning of the concatenation $P_{c}$ of
$\hat{P}=\{S_{c1}, S_{c2}, \dots S_{ch}\}$ as shown in Figure
\ref{fig_exampleFragmentGeneration}. The size $|F|$ is the number of canonical
instructions in the fragment $F$, and it is a hyper-parameter in
\textsf{PetaDroid}. Our grid search for the best $|F|$ hyper-parameter result
$|F|=10k$ for the current version of \textsf{PetaDroid}. For a sequence $S_{ci}$,
the order of canonical instructions is preserved within a method. In other
words, we guarantee  the preservation of order inside the method sequence or
what we refer to as a \textit{micro-action}. However, no specific order is
assumed between methods' sequences or what we refer to as \textit{macro-action}
(or behavior). On the contrary, before we truncate $P_{c}$ into size $|F|$, we
apply random permutations on $\hat{P}$ to produce a random order in the
macro-behavior.  The randomization happens in every access, whether it is during
training or deployment phases. Each Android sample has $\frac{h!}{(h-k)!}$
possible permutations for the methods' sequences
$\hat{P}=\{S_{c1}, S_{c2}, \dots S_{ch}\}$, where $h$ is the number of methods'
sequence in a given Android app, and $k$ is the number of sampled sequences. The
concatenation of the sampled $k$ sequences must be greater than $|F|$.

The intuition behind fragment detection is the abstraction of Android apps
behavior into very small \textit{micro-actions}. We consider each method
canonical instruction sequence $S_{c}$ as possible \textit{micro-actions} for an
Android app. In a fragment, we keep the possible micro-actions intact and
discard the app flow graph. We argue that this will force pattern learning,
during the training, to focus on only micro-actions, which allows better
generalization. Fragment-based detection has many advantages in the context of
malware detection. First, it challenges the machine learning model and its
training process to learn dynamic patterns at every training epoch. It also
focuses the model on robust, distinctive patterns from a sample of random
micro-actions of methods. Second, we argue that our fragment-based detection
improves the overall resiliency of the malware detection model against
conventional obfuscation techniques and code transformation. Third, in the
deployment phase, \textsf{PetaDroid} infers the maliciousness of a given sample
by applying \textsf{PetaDroid} CNN on multiple fragments from a single sample
to obtain a detection decision. 

{\it These fragments are entirely different since we apply the randomization
technique for every fragment generation. Furthermore, each code fragment $F$
produces a different view angle of the Android app for the machine learning
model and brings new information by considering a new $S_{ci}$ list in the
truncation. Finally, the fragment detection happens on a single CNN model
transparently to the ensemble.}


\subsubsection{Inst2Vec Embedding}
\label{sec_appraochInst2VecEmbedding}

\textit{Inst2Vec} is based on \textit{word2vec} \cite{Mikolov2013Distributed}
technique to produce an embedding vector for each canonical instruction in our
sequences.  \textit{Inst2Vec} is trained on instruction sequences to learn
instruction semantics from the underlying contexts. This means that
\textit{Inst2Vec} learns a dense representation of a canonical instruction that
reflects the instruction co-occurrence and context. The produced embeddings
capture the semantics of instructions (interpreted by geometric distances).
Furthermore, embedding features show high code fingerprinting accuracy and
resiliency to common obfuscation techniques \cite{Ding2019Asm2Vec}. Word2vec
\cite{Mikolov2013Distributed} is a vector space model to represent the words of
a document in a continuous vector space where words with similar semantics are
mapped closely in the space.  \textit{word2vec} functionality could be replaced
with recent representation learning models based on {\it Transformer} such as
\cite{DBLP:conf/naacl/PetersNIGCLZ18}, \cite{DBLP:conf/naacl/DevlinCLT19}.
From a security perspective, we want to map our features (canonical
instructions in a code fragment) to continuous vectors where their semantics is
translated to a distance in the vector space
\cite{DBLP:journals/pacmpl/AlonZLY19}.

Word2vec is a neural probabilistic model that is trained using the maximum
likelihood concept.  More precisely, given sequence of words: ${w_{1}, w_{2},
\cdots, w_{T}}$, at each position $t=1, \cdots, T$, the model predicts a
context of sequence within a window of fixed size $m$ given center word $w_j$
(illustrated in Equation \ref{equ_maximum_likelihood_du}), where $m$ is the
size of the training context \cite{Mikolov2013Distributed}.
\begin{align}
    L(\theta) &= \prod_{t=1}^{T}
                 \prod_{-m \leq j \leq +m, j \neq 0}
                 P(w_{t+j}| w_t; \theta)
    \label{equ_maximum_likelihood_du}
\end{align}
The objective function \cite{Mikolov2013Distributed} $J(\theta)$ is the
negative log likelihood, as shown in Equation \ref{equ_objective_function_du}.
The probability $P(w_{t+j}|w_t;\theta)$ is defined in Equation
\ref{equ_softmax}, where $v_w$ and $v'_w$ are the input and the output of the
embeddings of $w$.
\begin{align}
    J(\theta) &= - \frac{1}{T} \log L(\theta) \\
              &= - \frac{1}{T} \sum_{t=1}^{T}
              \sum_{-m \leq j \leq m, j \neq 0} \log P(w_{t+j} | w_t; \theta)
    \label{equ_objective_function_du}
\end{align}
\begin{align}
P(w_O | w_I) &= \text{softmax} (\grave{v}_{w_O}^T v_{w_I}) \\
             &= \frac{\exp ( \grave{v}_{w_O}^T v_{w_I} ) }
                     {\sum_{w=1}^W \exp ( \grave{v}_{w}^T v_{w_I} ) }
    \label{equ_softmax}
\end{align}
We train the embedding model by maximizing log-likelihood as illustrated in
Equation \ref{equ_log_likelihood_training_du}.
\begin{align}
     J_\text{ML} &= \log P(w_O | w_I) \\
                 &= (\grave{v}_{w_O}^T v_{w_I}) -
                 \log \left( \sum_{w=1}^W
                     \exp \{ \grave{v}_{w}^T v_{w_I} \} \right).
    \label{equ_log_likelihood_training_du}
\end{align}

\subsubsection{Classification Model}
\label{sec_appraochClassificationModel}

Our single CNN model takes \textit{Inst2Vec} features, which are a sequence of
embeddings; each embedding captures the semantics of an instruction. The
temporal CNN \cite{Kim2014Convolutional}, or 1-dimensional CNN
\cite{Xiang2015Character}, is the working core component in the {\sf PetaDroid}
single classification model. Table \ref{tab_singleCNNModel} details the
architecture of our CNN single model.

\begin{table}[!h]
\centering
\caption{PetaDroid Single CNN Model}
\label{tab_singleCNNModel}
\begin{scriptsize}
\begin{threeparttable}
\begin{tabular}{c|lp{4.8cm}}
     \hline
     \#&  \textbf{\textit{Layers}}    & \textbf{\textit{Options}}           \\
     \hline
     \hline
     1           & 1D-Conv            & Filter=128, Kernel=(5,5), Stride=(1,1), Padding=0, Activation=ReLU \\
     2           & BNorm              & BatchNormalization                  \\
     3           & Global Max Pooling & /                                   \\
    \hline\hline
     4           & Linear             & \#Output=512 , Activation=ReLU      \\
     5           & Linear             & \#Output=256 , Activation=ReLU      \\
     6           & Linear             & \#Output=1   , Activation=ReLU      \\
     \hline
\end{tabular}
\end{threeparttable}
\end{scriptsize}
\end{table}

We choose to build our classification models based on CNN architecture over
recurrent neural networks (RNN) such as LSTM or GRU. Due to the efficiency
of CNN during the training and the deployment compared to RNN
\cite{Goodfellow-et-al-2016}.
{\bf In the training phase}, the CNN models take on average 0.05 second per
batch (32 samples), which is five times faster than RNN models in our
experiments. The CNN model converges early (starting from 10 epochs) compared to
the RNN model (starting from 30 epochs). {\bf In the deployment phase}, the CNN
model's inference is, on average, five times faster than RNN models.
Both neural network architecture gives very similar detection results in our
experiments.  However, our automatic adaptation technique will benefit from the
efficiency of CNN models to rapidly build new models using large datasets. The
non-linearity used in our model employ the rectified linear unit (ReLUs)
\(h(x) = \max\{0, x\}\). We used Adam \cite{Goodfellow-et-al-2016} optimization
algorithm with a $32$ mini-batch size and  a \(3e-4\) learning rate  for $100$
epochs in all our experiments. The chosen hyper-parameters are the results of
empirical evaluations to find the best values.

\subsubsection{Dataset Notation}
\label{sec_appraochDatasetNotation}

In this section, we present the notations used in the next sections.

$X= \{\langle P_0, y_0 \rangle, \langle P_1, y_1 \rangle, .., \langle
P_{m}, y_{m} \rangle \}$: $X$ is the global dataset used to build ensemble
models and report \textsf{PetaDroid} performance on various tasks, where $m$ is
the number of $\langle sample, label \rangle$ records in the global dataset
$X$.

$X = \{X_{build}, X_{test}\}$: We use a build set $X_{build}$ to train and tune
the hyper-parameters of \textsf{PetaDroid} models. The test set $X_{test}$
represents Android apps that the system will receive during the deployment. The
test set $X_{test}$ is used to measure the final performance of
\textsf{PetaDroid}, which is reported in the evaluation section.  $X$ is split
randomly into $X_{build}$ ($50\%$) and $X_{test}$ ($50\%$).

$X_{build} = \{X_{train}, X_{valid}\}$: The build set, $X_{build}$, is composed
of a training set $X_{train}$ and a validation set $X_{valid}$. It is used to
build \textsf{PetaDroid} single CNN models for the CNN ensemble. For each
single CNN model, we tune the model parameters to achieve the best detection
performance on $X_{valid}$. The build set $m_{build} = m_{train} + m_{valid}:$
is the total number of records used to build \textsf{PetaDroid}. The training
set takes $80\%$ of the build set $X_{build}$ and, $20\%$ of $X_{build}$ is
used for the validation set $X_{valid}$.

\subsubsection{Detection Ensemble}
\label{sec_appraochDetectionEnsemble}

\textsf{PetaDroid} detection component relies on an ensemble $\Phi=\{
CNNModel_1, CNNModel_2, \dots CNNModel_{\phi}
\}$.
Ensemble $\Phi$ is composed of $\phi$
single CNN models.  The number of single CNN models in the ensemble $\phi$ is a
hyper-parameter. We choose to be $\phi=6$, which is a trade-off of between
maximum effectiveness on malware detection with the highest efficiency possible
base on our evaluation experiments. 

As mentioned previously, \textsf{PetaDroid} trains each CNN model for the
number of epochs ($epochs=100$). In each epoch, we compute $Loss_T$ and
$Loss_V$, the \textit{training} and \textit{validation} losses, respectively,
and save a snapshot of the single CNN model parameters. $Loss_T$ and $Loss_V$
are the log loss across training and validation sets:
\[
 p = CNNModel_{\theta}(y=1|F)
\]
\[
  loss(y, p) = -(y \log(p) + (1-y) \log(1-p)), 
\]
\[
  Loss_T  = \frac{-1}{m_{train}} \sum_{i=1}^{m_{train}} loss(y_i, p_i),
\]
\[
  Loss_V  = \frac{-1}{m_{valid}} \sum_{i=1}^{m_{valid}} loss(y_i, p_i),
\]
Where $p$ is the maliciousness likelihood probability given a fragment $F$ (a
truncated concatenation of canonical instructions $P_{c}$) and model parameters
$\theta$ (Section \ref{sec_appraochAppRepresentation}). \textsf{PetaDroid}
selects the top $\phi$ models automatically from the saved model snapshots that
have the lowest \textit{training} and \textit{validation} losses $Loss_T$ and
$Loss_R$, respectively.

\begin{equation}
\begin{small}
\begin{aligned}
  \hat{y} &= \Phi(x)
          &= \frac{1}{\phi} \left(\sum^{\phi}_{i} CNNModel_{i}(x) \right) \\
\end{aligned}
\end{small}
\label{equ_PetaDroidEnsembleDecision}
\end{equation}

\textsf{PetaDroid} CNN ensemble $\Phi$ produces a maliciousness probability
likelihood by averaging the likelihood probabilities of multiple CNN models,
as shown in Equation \ref{equ_PetaDroidEnsembleDecision}. 

\subsubsection{Confidence Analysis}
\label{sec_appraochConfidenceAnalysis}

\textsf{PetaDroid} ensemble computes the maliciousness probability likelihood
$Prob_{Mal}$ given a fragment $F$, as follows:
\[
  \hat{y}    = \Phi(F),~~
  Prob_{Mal} = \hat{y},~~
  Prob_{Ben} = (1 - \hat{y})
\]
Previous Android malware detection solutions, such as
\cite{mariconti2019mamadroid}, \cite{karbab2018maldozer},
\cite{Sen2016StormDroid}, utilize a simple detection technique (we refer to it
as a \textit{general decision}) to decide on the maliciousness of Android apps.
In the \textit{general decision}, we compute the general threshold $\zeta \in
[0, 1]$ that achieves the highest detection performance on the validation
dataset $X_{valid}$. In the deployment phase (or evaluation in our case on
$X_{test}$), The general decision $D_{\zeta}$ utilize the computed threshold
$\zeta$ to make detection decisions:
\[ D_{\zeta} =
  \begin{cases}
    Malware   & Prob_{Mal} >  \zeta\\
    Benign    & Prob_{Mal} <= \zeta\\
  \end{cases}
\]

\textsf{PetaDroid} employs f1-score as a detection performance metric to
automatically select $\zeta$ and to report the general detection performance on
the test set $X_{test}$ during our evaluation, in Section
\ref{sec_petaEvaluation}. We choose f1-score as our detection performance
metric due to its simplicity, and its measurement reflects the reality under
unbalanced datasets. {\it The general decision} provides a firm decision for
every sample. However, security practitioners might prefer dealing with
decisions that have associated confidence values and filter out less-confident
decisions for further investigations. In a real deployment, we want as many
detection decisions with high confidence and filter out the few uncertain apps
with low confidence probability. Unfortunately, the {\it general decision}
strategy that has been used by most previous solutions does not provide such
functionality. For this purpose, we propose the \textbf{confidence decision
strategy}, a mechanism to automatically filter out apps with uncertain
decisions.  \textsf{PetaDroid} computes a confidence threshold $\eta$ that
achieves a high detection performance (f1-score) and a negligible error rate
(false negative and false positive rates) in the validation dataset. In other
words, we add the error rate constraint to the system that computes the
detection threshold $\eta$ from $X_{valid}$. In the deployment, we make the
confidence-based decision as follow:

\begin{scriptsize}
\[ D_{\eta} =
  \begin{cases}
    Uncertain & Prob_{Mal}  < \eta~\land~Prob_{Ben} < \eta      \\
    Malware   & Prob_{Mal} >= \eta~\land~Prob_{Mal} > Prob_{Ben}\\
    Benign    & Prob_{Ben} >= \eta~\land~Prob_{Ben} > Prob_{Mal}\\
  \end{cases}
\]
\end{scriptsize}
For example, we could fix the error rate to $<1\%$ and automatically find
$\eta$ that achieves the highest f1-score in the validation set. Our goal is to
maximize confidence detection decisions during the deployment, which we called
the \textit{detection coverage performance} and minimize alerts for uncertain
ones that require further analyses (such as dynamic analyses).  In our case,
the \textit{detection coverage performance} is the percentage of confidence
decisions from $X_{test}$. In Section \ref{sec_petaEvaluation}, we conduct
experiments where we report the \textit{general detection performance} metric
to compare with existing solutions such as \cite{mariconti2019mamadroid},
\cite{karbab2018maldozer}, \cite{Sen2016StormDroid}. Besides, we report
\textit{confidence detection performance} and \textit{detection coverage
performance} metrics, which we believe are suitable for real-world deployment.
The \textit{confidence decision strategy} is key for automatic adaptation in
\textsf{PetaDroid}, as will be explained next.

\subsubsection{Automatic Adaptation}
\label{sec_appraochExtensionUpdate}

In this section, we describe our mechanism to adapt to Android ecosystem
changes over time automatically. The key idea is to re-train the CNN ensemble
on new benign and malware samples periodically to learn the latest changes. To
enhance the automatic adaptation, we leverage the confidence analysis to
collect an extension dataset that captures the incremental change over time.
Initially, we train \textsf{PetaDroid} ensemble using $X_{build}=\{X_{train} +
X_{valid}\}$. Afterward, \textsf{PetaDroid} leverages the \textit{confidence
detection strategy} to build an extension dataset $X_{exten}$ from test dataset
$X_{test}$ from high-confidence detected apps. In a real deployment, $X_{test}$
is a stream of Android apps that needs to be checked for maliciousness by the
vetting system. The test dataset $X_{test}=\{X_{Certain}, X_{Uncertain}\}$ is
composed of apps having a high-confidence decision ($X_{Certain}$ or
$X_{exten}$) and apps having uncertain decisions $X_{Uncertain}$.  In the
deployment, \textsf{PetaDroid} accumulates from high-confidence apps over time
to form $X_{exten}$ dataset. Periodically, \textsf{PetaDroid} utilizes the
extension dataset $X_{exten}$ to extend the original $X_{build}$ and later
updates the CNN ensemble models. In our evaluation, and after updating the CNN
ensemble, we report \textbf{updated general performance} and \textbf{updated
confidence-based performance}, respectively the general and confidence-based
performance of the new trained CNN ensemble on $X_{test}$. These metrics answer
the question: what would be the detection performance on
$X_{test}=\{X_{Certain}, X_{Uncertain}\}$ after we build the ensemble on
$X_{NewBuild}=\{X_{Certain}, X_{build}\}$? In other words, \textsf{PetaDroid}
reviews previous detection decisions using the new CNN ensemble and drives new
general and confidence-based performance.

In a deployment environment, \textsf{PetaDroid} is continuously receiving new
Android apps, whether benign or malware, which is figuratively our $X_{test}$.
\textsf{PetaDroid} employs the extension dataset $X_{exten}$ to automatically
overcome pattern changes, whether malicious or benign. Our approach is based on
the assumption that Android apps patterns change incrementally with slow
progress. Therefore, starting from a relatively small $X_{build}$
dataset,\textsf{PetaDroid} could learn new patterns from new $X_{exten}$
datasets progressively over time. \textsf{PetaDroid} ensemble update is an
automatic operation for every period. Our evaluation (Section
\ref{sec_petaAuto}) shows the effectiveness of our automatic adaptation
strategy.

\subsection{Malware Clustering}
\label{sec_appraochClustering}

In this section, we detail the family clustering system.
\textsf{PetaDroid} clustering aims to group the previously detected malicious
apps (Section \ref{sec_appraochMalwareDetection}) into highly similar malicious
apps groups, which are most likely part of the same malware family.
\textsf{PetaDroid} clustering process starts from a multiset of discretized
canonical instruction sequences $P=\{S_{c1}, S_{c2}, \dots S_{ch}\}$ of
the detected malicious apps. We introduce the \textit{InstNGram2Vec} technique
and deep neural network auto-encoder to generate embedding digests for
malicious apps. Afterward, we cluster malware digests using the DBScan
\cite{DBLP:conf/kdd/EsterKSX96} clustering algorithm to generate malware family
groups.
\subsubsection{InstNGram2Vec}
\label{sec_appraochInstNGram2Vec}
Notice that our clustering system (DBScan \cite{DBLP:conf/kdd/EsterKSX96})
requires to represent malware samples by one feature vector for each sample
instead of a list of embeddings as in {\it Inst2Vec} for {\sf PetaDroid}
classification. For this reason, we introduce {\it InstNGram2Vec} technique
that automatically represents malware samples as feature vectors without an
explicit manual feature selection. \textit{InstNGram2Vec} is a technique that
maps concatenated instruction sequences to fixed-size embeddings employing NLP
bag of words (N-grams) \cite{ngram2004AbouAssaleh} and feature hashing
\cite{qinfeng09hashk} techniques.

\paragraph{Common N-Gram Analysis (CNG)}
\label{sec_ngrams}
The common N-gram analysis (CNG) \cite{ngram2004AbouAssaleh}, or simply N-gram,
has been extensively used in text analyses and natural language processing in
general and related applications such as automatic text classification and
authorship attribution \cite{ngram2004AbouAssaleh}. N-gram computes the
contiguous sequences of $n$ items from a large sequence. In the context of
\textsf{PetaDroid}, we compute canonical instructions N-grams on concatenated
sequence $P_{c}$ by counting the instruction sequences of size $n$. Notice that
the N-grams are extracted using a forward-moving window (of size $n$) by one
step and incrementing the counter of the found features (instruction sequence
in the window) by one. The window size $n$ is a hyper-parameter; we choose
$n=4$ based on our experimentations. We notice that using $n>4$ will affect the
efficiency considerably if feature vector generation. Using $n<4$ affects the
effectiveness of the clustering. N-gram computation takes place simultaneously
with the feature hashing in the form of a pipeline to prevent and limit
computation and memory overuse due to the high dimensionality of N-grams.

\paragraph{Feature Hashing}
\label{sec_feature_hashing}
\textsf{PetaDroid} employs Feature Hashing (FH) \cite{qinfeng09hashk} along
with N-grams to vectorize $P_{c}$.  The feature hashing algorithm takes as an
input $P_{c}$ N-grams generator and the target length $L$ of the feature
vector. The output is a feature vector with components $x_i$ and a fixed size
$L$. In our framework, we fix $L=|V|$, where $V$ is the vocabulary dictionary
(Section \ref{sec_appraochAppRepresentation}). We normalize $x_i$ using the
euclidean norm (also L2 norm). Applying \textit{InstNGram2Vec} on a detected
malicious app $P_{c}$ produces a fixed size hashing vector $hv$. Therefore, the
result is $HV=\{hv_0, hv_1, \dots hv_{DMal}\}$, and hashing vector $hv$ for
$DMal$ detected malicious apps.
\begin{equation} \label{equ_l2norm}
L2Norm(x) =  \|x\|_2 = \sqrt{x^{2}_{1} +.. +  x^{2}_{n}}
\end{equation}
The feature hashing algorithm takes as an input $P_{c}$ N-grams generator and
the target length $L$ of the feature vector. The output is a feature vector
$x_i$ with a fixed size of $L$. In our framework, we fixed $L=|V|$, where $V$
is the vocabulary dictionary (Section \ref{sec_appraochAppRepresentation}) to
prevent collusion problems. We normalize $x_i$ using the euclidean norm. As
shown in Formula \ref{equ_l2norm}, the euclidean norm is the square root of the
sum of the squared vector values. Previous researches
\cite{Weinbergeretal09,qinfeng09hashk} shows that the hash kernel approximately
preserves the vector distance and grows linearly with the number of samples.
Applying \textit{InstNGram2Vec} on a detected malicious app $P_{c}$ produces a
fixed size hashing vector $hv$. Therefore, the result is $HV=\{hv_0, hv_1,
\dots hv_{DMal}\}$, a hashing vector $hv$ for $DMal$ detected malicious apps.

\subsubsection{Auto-Encoder}
\label{sec_appraochAutoEncoder}

We develop a deep neural auto-encoder through stacked neural layers of encoding
and decoding operations, as shown in Table \ref{tab_autoEncoderArch}. The
proposed auto-encoder learns the latent representation of Android apps in an
unsupervised way. The unsupervised learning of the auto-encoder is done through
the reconstruction (Table \ref{tab_autoEncoderArch}) of the unlabeled hashing
vectors $HV=\{hv_0, hv_1, \dots hv_{DMal}\}$ of random Android apps. Notice
that we do not need any labeling during the training of \textsf{PetaDroid}
auto-encoder, off-the-self Android apps are sufficient.

\begin{table}[!h]
\centering
\caption{PetaDroid Neural Auto-Encoder}
\label{tab_autoEncoderArch}
\begin{scriptsize}
\begin{threeparttable}
    \begin{tabular}{c|ll}
     \hline
     \#&  \textbf{\textit{Layers}}     & \textbf{\textit{Options}} \\
     \hline
     \hline
     01 & Linear & \#Output=$|V|$, Activation=Tanh \\
     02 & Linear & \#Output=512  , Activation=Tanh \\
     03 & Linear & \#Output=256  , Activation=Tanh \\
     04 & Linear & \#Output=128  , Activation=Tanh \\
     05 & Linear & \#Output=64   , Activation=Tanh \\
     \hline
     06 & Linear & \#Output=64   , Activation=Tanh \\
     07 & Linear & \#Output=128  , Activation=Tanh \\
     08 & Linear & \#Output=256  , Activation=Tanh \\
     08 & Linear & \#Output=512  , Activation=Tanh \\
     10 & Linear & \#Output=$|V|$, Activation=Tanh \\
     \hline
\end{tabular}
\end{threeparttable}
\end{scriptsize}
\end{table}

The training goal is to make the auto-encoder learn to efficiently produce a
latent representation (or digest) of an Android app $hv$ that keeps the
discriminative patterns of malicious and benign Android apps. Formally, the
input to the deep neural auto-encoder \cite{DBLP:conf/icann/HintonKW11} network
is an unlabeled hash vector $HV=\{hv_0, hv_1, \dots hv_{DMal}\}$, denoted
$\bm{x^{\prime}} \in \mathcal{U}$ on which operates the encoder network
$f_{\text{enc}}: \mathbb{R}^{|V|}\to\mathbb{R}^{p}$, $p=64$ as shown in Table
\ref{tab_autoEncoderArch} (parameterized by $\Theta_{\text{enc}}$) to produce
the latent representation $\bm{z}_{\bm{x}^{\prime},\Theta_{\text{enc}}}$,
\emph{i.e.}

\begin{equation}
\bm{z}_{\bm{x}^{\prime},\Theta_{\text{enc}}} =
  f_{\text{enc}}(\bm{x^{\prime}};\Theta_{\text{enc}})
\label{eq:z}
\end{equation}

The produced digest, namely $\bm{z}_{\bm{x}^{\prime}, \Theta_{\text{enc}}} \in
\mathbb{R}^{p}$, is used by the decoder network $f_{\text{dec}}: \mathbb{R}^{p}
\to \mathbb{R}^{|V|}$ to rebuild or reconstruct the {\it InstNGramBag2Vec}
feature vector. The training loss of the auto-encoder network given the
unlabeled $hv$ $\bm{x^{\prime}}$ is,

\begin{equation}
\bm{\tilde{x}^{\prime}} = f_{\text{dec}}(\bm{z};\Theta_{\text{dec}})
\label{eq:x_recon}
\end{equation}

$\bm{\tilde{x}^{\prime}}\in \mathbb{R}^{d\times w} $ denotes the generated
reconstruction.

\begin{equation}
\mathcal{L}_{\text{auto}}(\bm{x^{\prime}};
  \Theta_{\text{enc}},\Theta_{\text{dec}})= {\lVert \bm{x^{\prime}}-
  f_{\text{dec}}(\bm{z}_{\bm{x}^{\prime},
  \Theta_{\text{enc}}};\Theta_{\text{dec})} \rVert}^{2}
\label{eq:L_auto}
\end{equation}

In the training phase, the gradient-based optimizer minimizes the objective
reconstruction function on the {\it InstNGramBag2Vec} feature vectors of
unlabeled Android apps.

\begin{equation}
(\Theta^{*}_{\text{enc}}, \Theta^{*}_{\text{dec}}) = \arg \min_{
  \Theta_{\text{enc} }, \Theta_{\text{dec}}} \sum_{i=1}^{N_1+N_2}
\mathcal{L}_{\text{auto}} ( \bm{x^{\prime}}_i; \Theta_{\text{enc}},
  \Theta_{\text{dec}})
\label{eq:opt_solauto}
\end{equation}

\textit{Notice that PetaDroid auto-encode is trained only once during all the
experimentation due to its general usage.} To this end, \textsf{PetaDroid}
employs a trained encoder $f_{\text{dec}}$ to produce digests $\bm{Z} =
\{\bm{z_0}, \bm{z_1}, \dots, \bm{z_{DMal}}\}$  for the detected malicious apps.

\subsubsection{Family Clustering}
\label{sec_approachClustering}

\textsf{PetaDroid} clusters the detected malware digests $\bm{Z}=\{\bm{z_0},
$ $\bm{z_1}, \dots, \bm{z_{DMal}}\}$ into groups of malware with high similarity
and most likely belonging to the same family. In \textsf{PetaDroid} clustering:
{\bf First}, we use an \textbf{exclusive} clustering mechanism. The clustering
algorithm only groups highly similar samples and tags the rest as
non-clustered. This feature could be more convenient for real-world deployments
since we might not always detect malicious apps from the same family, and we
would like to have family groups only if there are groups of the sample malware
family. To achieve this feature, we employ the \textit{DBScan} clustering
algorithm.  {\bf Second}, as an optional step, we find the best cluster for the
non-cluster samples, from the clusters produced previously by computing the
euclidean similarity between a given non-cluster sample and a given cluster
samples. We call this step the {\it family matching}, as shown in Figure
\ref{fig_petadroid_approach}.  In the evaluation, we report {\it homogeneity}
and {\it coverage} metric for the clustering before and after applying this
optional step. \textit{DBScan}, in contrast with clustering algorithms such as
\textit{K-means}, produces clusters with high confidence. The most important
metrics in \textsf{PetaDroid} clustering is the homogeneity of the produces
clusters.

\section{Implementation}
\label{sec_implementation}

We build \textsf{PetaDroid} using \texttt{Python} and \texttt{Bash} programming
languages.  We use \texttt{dexdump}\footnote{https://tinyurl.com/y4ze8nyy} to
disassemble the DEX bytecode into Dalvik assembly. The tool \texttt{dexdump} is
a simple and yet very efficient tool to parse APK file and produce disassembly
in a textual form. We develop python and bash scripts to parse Dalvik assembly
to produce sequences of canonical instructions. Notice that there is no
optimization in the preprocessing; in the efficiency evaluation, we only use a
single thread script for a given Android app. We implement \textsf{PetaDroid}
neural networks, CNN ensemble, and auto-encoders, using
PyTorch\footnote{https://pytorch.org}.  For clustering, we employ official
\texttt{hdbscan}\footnote{https://hdbscan.readthedocs.io} implementation.  We
evaluate the efficiency of \textsf{PetaDroid} on a commodity hardware server
(Intel(R) Xeon(R) CPU E5-2630, 2.6GHz). For training, we use \textit{NVIDIA
8 x Titan RTX} Graphic Processing Unit (GPU).

\section{Dataset}
\label{sec_dataset}

Our evaluation dataset contains 10 million Android apps as sampling space for
our experiments (over 100TB) collected across the last ten years from August
2010 to August 2019, as depicted in Table \ref{tab_datasetSummary}. The
extensive coverage in size (10 M), time range (06-2010 to 08-2019), and malware
families (+300 family) make the result of our evaluation quite compelling.

In section \ref{sec_petaMalwareDetetion} and \ref{sec_familyClustering}, to
evaluate \textsf{PetaDroid} detection and family clustering, we leverage malware
from reference Android malware datasets, namely: MalGenome
\cite{zhou2012dissecting}, Drebin \cite{arp2014drebin}, MalDozer
\cite{karbab2018maldozer}, and AMD \cite{wei2017deep}. Also, we collected
Android malware from VirusShare \footnote{https://VirusShare.com} malware
repository.  In addition, we use benign apps from AndroZoo
\cite{Allix2016AndroZoo} dataset (randomly sampling from 7.4 Million benign
samples in each experiment). In the family clustering evaluation (section
\ref{sec_familyClustering}), we use only malware samples from the reference
datasets.

\begin{table}[h]
\centering
\caption{Evaluation Datasets}
\label{tab_datasetSummary}
\begin{scriptsize}
\centering
\begin{tabular}{l c c c}
\hline
\textbf{Name} & \textbf{\#Samples} & \textbf{\#Families} & \textbf{Time} \\
\hline
\texttt{MalGenome}  \cite{zhou2012dissecting}           & 1.3K & 49  & 2010-2011      \\
\texttt{Drebin}     \cite{arp2014drebin}                & 5.5k & 179 & 2010-2012      \\
\texttt{MalDozer}   \cite{karbab2018maldozer}           & 21k  & 20  & 2010-2016      \\
\texttt{AMD}        \cite{wei2017deep}                  & 25k  & 71  & 2010-2016      \\
\texttt{VirusShare} \footnote{https://VirusShare.com}   & 33k  &  /  & 2010-2017      \\
\texttt{MaMaDroid}  \cite{mariconti2019mamadroid}       & 40k  &  /  & 2010-2017      \\
\texttt{AndroZoo}   \cite{Allix2016AndroZoo}            & 9.5M &  /  & 2010- Aug 2019 \\
\hline
\end{tabular}
\end{scriptsize}
\end{table}

In the comparison (Section \ref{sec_compStudy}) between \textsf{PetaDroid},
\textsf{MaMaDroid} \cite{mariconti2017mamadroid,mariconti2019mamadroid}, and
\textsf{DroidAPIMiner} \cite{Aafer2013DroidAPIMiner}, we apply
\textsf{PetaDroid} on the same dataset (benign and malware) used in MaMaDroid
evaluation\footnote{https://bitbucket.org/gianluca\_students/mamadroid\_code/src/master/}
to measure the performance of \textsf{PetaDroid} against state-of-the-art
Android malware detection solutions.

\begin{figure}[t]
\centering
\includegraphics
[width=0.45\textwidth, trim=0cm 0cm 0cm 0cm,clip]
{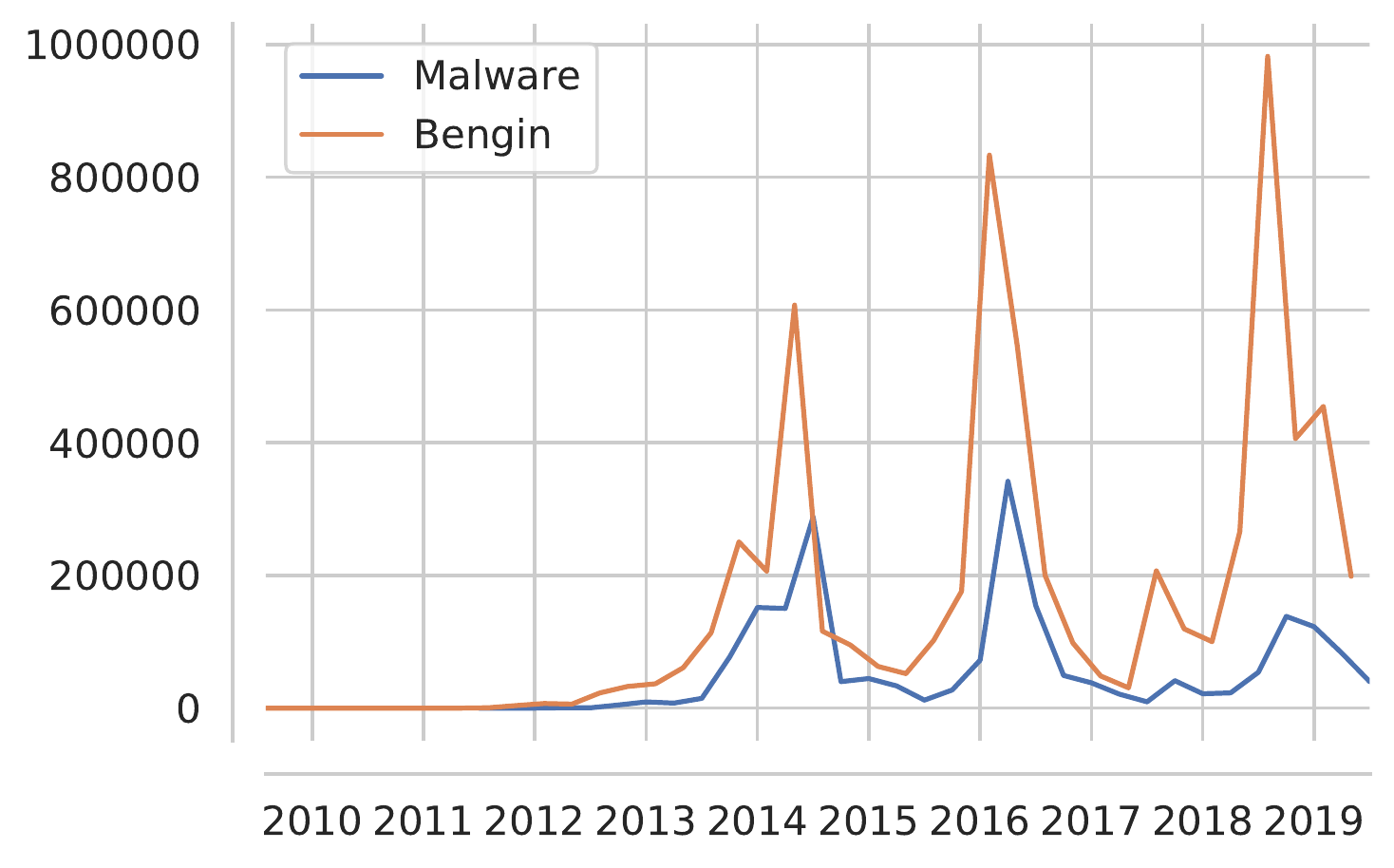}
\caption{PetaDroid Dataset Distribution Over Time}
\label{fig_androzooDistTime}
\end{figure}

To assess \textsf{PetaDroid} obfuscation resiliency (Section
\ref{sec_petaObfusResiliency}), we conduct an obfuscation evaluation on PRAGuard
dataset\footnote{http://pralab.diee.unica.it/en/AndroidPRAGuardDataset}, which
contains $11k$ obfuscated malicious apps using common obfuscation techniques
\cite{Davide2015Praguard}. Besides, we generate over $100k$ benign and malware
obfuscated Android apps employing DroidChameleon obfuscation tool
\cite{Vaibhav2013DroidChameleon} using common obfuscation techniques and their
combinations.

To assess the adaptation of {\sf PetaDroid} (Section \ref{sec_petaAuto}), we
employ the whole
AndroZoo\footnote{https://androzoo.uni.lu/}~\cite{Allix2016AndroZoo} dataset
(until August 2019), which contains 7.4 million benign apps and 2.1 million
malware apps, by randomly sampling a dataset ($100k$ malware and bengin) in each
experiment. We rely on VirusTotal detection of multiple anti-malware vendors in
(metadata provided by AndroZoo repository) to label the samples.  The dataset
covers more than ten years span of Android benign and malware apps
~\cite{Allix2016AndroZoo}.


\section{Evaluation} 
\label{sec_petaEvaluation}

In this section, we evaluate \textsf{PetaDroid} framework through a set of
experiments and settings involving different datasets.

We aim to answer questions such as: {\it What is the detection performance of
\textsf{PetaDroid} on small and large training datasets (Section
\ref{sec_petaMalwareDetetion})?} {\it What is the effect of \textsf{PetaDroid}
ensemble and build dataset sizes on the overall performance (Section
\ref{sec_petaDatasetSizeEffect})?} {\it What is the performance of family
clustering (Section \ref{sec_familyClustering})?} {\it How efficient is
\textsf{PetaDroid} in terms of runtime on commodity hardware  (Section
\ref{sec_petaEfficiency})?} {\it How robust is PetaDroid against common
obfuscation techniques (Section \ref{sec_petaObfusResiliency})?}

\subsection{Evaluation Metrics}
\label{sec_evalutionMetrics}

The evaluation results are presented in terms of \textit{precision},
\textit{recall}, and \textit{f1 score}.  We use \textit{homogeneity}
\cite{Andrew2007Measure} and \textit{coverage} metrics to measure the family
clustering performance. The homogeneity metric scores the purity of the
produced family clusters. A perfect homogeneity means each produced cluster
contains samples from only one malware family since \textsf{PetaDroid}
clustering aims only to generate groups with confidence while ignoring less
certain groups. The coverage metrics score the percentage of the clustered
dataset with confidence.

\textit{Precision} (P) is the percentage of positive prediction, i.e., the
percentage  of the detected malware out of all sample apps; formally, $P =
\frac{TP}{TP + FP}$.  \textit{Recall} (R). It is the percentage of correctly
detected malicious apps out of all malware samples, formally, $R =
\frac{TP}{TP + FN}$. \textit{True positives} (TP) measures the number of
correctly detected malicious apps.  \textit{False negatives} (FN): measures
the number of incorrectly classified malicious apps. \textit{False positives}
(FP):  measures the number of incorrectly classified benign apps.
\textit{f1-Score} (F1) is the harmonic mean of precision and recall, formally,
$f1= 2 ~\textsf{x}~ \frac{P \textsf{x} R}{P + R}$.

\subsection{Malware Detection}
\label{sec_petaMalwareDetetion}

In this section, we report the detection performance of \textsf{PetaDroid} and
the effect of hyper-parameters on malware detection performance.

\subsubsection{Detection Performance}
\label{sec_petaMalwareDetetionPerformance}

Table \ref{tab_detectionPerformance} shows \textsf{PetaDroid} \textit{general}
and \textit{confidence-based} performance in terms of f1-score, recall, and
precision metrics on the reference datasets. In the general performance,
\textsf{PetaDroid} achieves a high f1-score $96-99\%$ with a low false-positive
rate (precision score of $96.4-99.5\%$ in the general detection). The detection performance is higher under
confidence settings. The f1-score is $99\%$ and a very low false-positive rate
with a recall score of $99.8\%$ on average. The confidence-based performance
causes the filtration of $1-8\%$ low confidence samples from the testing set.
In all our experiments, the confidence performance flags $\approx6\%$ on
average, as uncertain decisions, which is a small and realistic value in a
deployment with low false positives. 



\begin{table}[ht!]
\centering
\caption{General and Confidence Performances on Various Reference Datasets}
\label{tab_detectionPerformance}
\begin{scriptsize}
\begin{tabular}{l|c|c}
\hline
\textbf{Name}       &  \textbf{General (\%)} & \textbf{Confidence (\%)} \\
                    & F1~~~-~~~~P~~~-~~~~R   & F1~~~-~~~~~P~~-~~~R      \\
\hline
\texttt{Genome}     & 99.1~~-~~99.5~~-~~98.6 & 99.5~~-~~100.~~-~~99.0   \\
\texttt{Drebin}     & 99.1~~-~~99.0~~-~~99.2 & 99.6~~-~~99.6~~-~~99.7   \\
\texttt{MalDozer}   & 98.6~~-~~99.0~~-~~98.2 & 99.5~~-~~99.7~~-~~99.4   \\
\texttt{AMD}        & 99.5~~-~~99.5~~-~~99.5 & 99.8~~-~~99.7~~-~~99.8   \\
\texttt{VShare}     & 96.1~~-~~96.4~~-~~95.7 & 99.1~~-~~99.7~~-~~98.6   \\
\hline
\end{tabular}
\end{scriptsize}
\end{table}


\subsubsection{Dataset Size Effect}
\label{sec_petaDatasetSizeEffect}

In Table \ref{tab_effectBuildSize}, there is a small change in the detection
performance when the build set percentage drops from $90\%$ to $50\%$ from the
overall dataset.  Note that the build dataset is already composed of $80\%$
training and $20\%$ validation set $X_{build} = \{X_{train}, X_{valid}\}$,
which makes the model trained on a smaller dataset. However, \textsf{PetaDroid}
detection still performs well under these settings. Notice that in all our
experiments, we use $50\%$ from the evaluation dataset as a build dataset.

\begin{table}[ht!]
\centering
\begin{scriptsize}
\begin{tabular}{l|c|c}
\hline
                                   &  \textbf{General (F1 \%)} & \textbf{Confidence (F1 \%)}\\
\textbf{Build Dataset Size (\%)}   & 50\%~-~70\%~-~90\% & ~50\%~-~70\%~-~90\%  \\
\hline
\texttt{Genome}                    & 98.8~-~99.1~-~98.8 & ~100.~-~99.5~-~99.1  \\
\texttt{Drebin}                    & 98.2~-~99.1~-~99.1 & ~99.6~-~99.6~-~99.8  \\
\texttt{MalDozer}                  & 98.3~-~98.6~-~98.7 & ~99.6~-~99.5~-~99.6  \\
\texttt{AMD}                       & 99.3~-~99.5~-~99.5 & ~99.7~-~99.8~-~99.7  \\
\texttt{VShare}                    & 95.6~-~96.1~-~96.4 & ~99.0~-~99.1~-~99.1  \\
\hline
\end{tabular}
\end{scriptsize}
\caption{Effect of Building Dataset Size on the Detection Performance}
\label{tab_effectBuildSize}
\end{table}
-------------------------------

\subsubsection{Ensemble Size Effect}
\label{sec_petaEnsemelbeSizeEffect}

Another important factor that affects \textsf{PetaDroid} malware detection
performance is the number of CNN models in the detection ensemble. Table
\ref{tab_effectEnsembleSize} depicts \textsf{PetaDroid} performance under
different ensemble sizes. We notice the high detection accuracy using a single
CNN model ($95-99\%$ f1-score). In addition to the strength of CNN in
discriminating Android malware, fragment detection adds a significant value to
the overall performance even in a single CNN model. In the case of {\tt
MalGenome} (Table \ref{tab_effectEnsembleSize}), the ensemble size adds no
value to the detection performance due to the small size of {\tt MalGenome}
dataset ($1.3k$ malware + $12k$ benign randomly sampled from AndroZoo
\cite{Allix2016AndroZoo}). In the case of {\tt VirusShare} (Table
\ref{tab_effectEnsembleSize}), augmenting the ensemble size enhanced the
detection rate. Our empirical tests show that $\phi=6$ as the ensemble size
gives the best detection results while keeping the system's efficiency.

\begin{table}[ht!]
\centering
\begin{scriptsize}
\begin{tabular}{l|c|c}
\hline
                   &  \textbf{General (F1 \%)} & \textbf{Confidence (F1 \%)}\\
\textbf{\#Model }  & 1~~-~~~5~~~~-~~10~~-~~20  & 1~~-~~~5~~~~-~~10~~-~~20   \\
\hline                                                                        
\texttt{MalGenome} & 99.5~-~99.3~-~99.3~-~99.1 & 99.5~-~99.5~-~99.5~-~99.5  \\
\texttt{Drebin}    & 99.0~-~99.1~-~99.0~-~99.1 & 99.4~-~99.6~-~99.6~-~99.6  \\
\texttt{MalDozer}  & 98.0~-~98.6~-~98.4~-~98.6 & 99.0~-~99.5~-~99.5~-~99.5  \\
\texttt{AMD}       & 99.3~-~99.5~-~99.5~-~99.5 & 99.5~-~99.8~-~99.7~-~99.8  \\
\texttt{VShare}    & 95.0~-~96.0~-~96.1~-~96.1 & 98.1~-~99.0~-~99.2~-~99.1  \\
\hline
\end{tabular}
\end{scriptsize}
\caption{Effect of Ensemble Size on Detection}
\label{tab_effectEnsembleSize}
\end{table}


\subsection{Family Clustering}
\label{sec_familyClustering}

In this section, we present the results of \textsf{PetaDroid} family clustering
on reference datasets (only malware apps). Malware family
clustering phase comes after \textsf{PetaDroid} detects a considerable number of
malicious Android apps. The number of detected apps could vary from $1k$
(MalGenome \cite{zhou2012dissecting}) to $+20k$ (Maldozer
\cite{karbab2018maldozer}) samples depending on the deployment. We use
\textit{homogeneity} \cite{Andrew2007Measure} and \textit{coverage} metrics to
measure the family clustering performance. The homogeneity metric scores the
purity of the produced family clusters. A perfect homogeneity means each
produced cluster contains samples from only one malware family. By default,
\textsf{PetaDroid} clustering aims only to generate groups with confidence-based
while ignoring less certain groups. The coverage metrics score the percentage of
the clustered dataset with confidence. We also report the clustering performance
after applying the {\it family matching} (optional step) to cluster all the
samples in the dataset ($100\%$ coverage).

\begin{table}[ht!]
\centering
\caption{The Performance of the Family Clustering}
\label{tab_clusteringResults}
\begin{scriptsize}
\begin{tabular}{l|c|c}
\hline
                          &  \textbf{DBSCAN Clustering}       & \textbf{After Family Matching}\\
\textbf{Clustering Metrics}         &~Homogeneity~|~Coverage~ &~Homogeneity~|~Coverage~\\
\hline
\texttt{Genome}                    &~~~90.00\%~|~37\%         &~~~79.67\%~~|~~100\%\\
\texttt{Drebin}                    &~~~92.28\%~|~49\%         &~~~80.48\%~~|~~100\%\\
\texttt{MalDozer}                  &~~~91.27\%~|~55\%         &~~~81.58\%~~|~~100\%\\
\texttt{AMD}                       &~~~96.55\%~|~50\%         &~~~81.37\%~~|~~100\%\\
\hline

\end{tabular}
\end{scriptsize}
\end{table}


Table \ref{tab_clusteringResults} summarizes the clustering performance in
terms of {\it homogeneity} and {\it coverage} scores before and after applying
the {\it family matching}. {\bf First}, \textsf{PetaDroid} can produce clusters
with high {\it homogeneity} $90-96\%$ while keeping an acceptable {\it
coverage}, $50\%$ on average. At first glance, $50\%$ {\it coverage} seems to
be a modest result, but we argue that it is satisfactory because: (i) we could
extend the coverage, but this might affect the quality of the produced
clusters. In the deployment, high confidence clusters with minimum errors and
acceptable coverage might be better than perfect coverage (in the case of
K-Means clustering algorithm) with a high error rate. (ii) The evaluation
datasets have long tail malware families, meaning that most families have only
a few samples.  This makes the clustering very difficult due to the few samples
(less than five samples) in each malware family in the detected dataset. In a
real deployment, we could add non-cluster samples to the next clustering
iterations. In this case, we might accumulate enough samples to cluster for the
long tail malware families. {\bf Second}, after applying the family matching,
{\sf PetaDroid} clusters all the samples in the dataset ($100$\% coverage) and
homogeneity decreased to $80-82\%$, which is acceptable.

\subsection{Obfuscation Resiliency}
\label{sec_petaObfusResiliency}

In this section, we report \textsf{PetaDroid} detection performance on
obfuscated Android apps. We experiment on: (1) PRAGuard obfuscation dataset
\cite{Davide2015Praguard} ($10k$) and (2) obfuscation dataset generated using
DroidChameleon \cite{Vaibhav2013DroidChameleon} obfuscation tool ($100k$).  In
the PRAGuard experiment, we combine PRAGuard dataset with $20k$ benign Android
apps randomly sampled from the benign apps of AndroZoo repository. We split the
dataset equally into build dataset $X_{build}=\{X_{train}, X_{valid}\}$ and
test dataset $X_{test}$. Table \ref{tab_praguardObfuscation} presents the
detection performance of \textsf{PetaDroid} on different obfuscation
techniques.  \textsf{PetaDroid} shows high resiliency to common obfuscation
techniques by having an almost perfect detection rate, $99.5\%$ f1-score on
average.

\begin{table}[ht!]
\centering
        \caption{PetaDroid Obfuscation Resiliency on PRAGuard Dataset (\textbf{General Performance})}
\label{tab_praguardObfuscation}
\begin{scriptsize}
\begin{tabular}{ll|ccc}
\hline
ID & \textbf{Obfuscation Techniques} & \textbf{F1 (\%)} & \textbf{P (\%)}                   & \textbf{R (\%)} \\
\hline
1 & {\tt Trivial              }  & 99.4 & 99.4 & 99.4 \\
2 & {\tt String Encryption    }  & 99.4 & 99.3 & 99.4 \\
3 & {\tt Reflection           }  & 99.5 & 99.5 & 99.5 \\
4 & {\tt Class Encryption     }  & 99.4 & 99.4 & 99.5 \\
5 & {\tt (1) + (2)            }  & 99.4 & 99.4 & 99.4 \\
6 & {\tt (1) + (2) + (3)      }  & 99.4 & 99.3 & 99.5 \\
7 & {\tt (1) + (2) + (3) + (4)}  & 99.5 & 99.4 & 99.6 \\
\hline
  & {\bf Overall}                & 99.5 & 99.6 & 99.4 \\
\hline
\end{tabular}
\end{scriptsize}
\end{table}

In the DroidChameleaon experiment, we evaluate \textsf{PetaDroid} on other
obfuscation techniques, as shown in Table \ref{tab_chameleonObfuscation}. The
generated dataset contains obfuscated benign ($5k$ apps randomly
sampled from AndroZoo) and malware samples (originally from Drebin). In the
building process of CNN ensemble, we only train with one obfuscation technique
(Table \ref{tab_chameleonObfuscation}) and make the evaluation on the rest of
the obfuscation techniques. Table \ref{tab_chameleonObfuscation} reports the
result of obfuscation resiliency on DroidChameleon generated dataset.  The
results show the robustness of PetaDroid. According to this experiment,
\textsf{PetaDroid} is able to detect malware obfuscated with common techniques
even if the training is done on non-obfuscated datasets. We believe that
\textsf{PetaDroid} obfuscation resiliency comes from the usage of (1) Android
API (canonical instructions) sequences as features in the machine learning
development. Android APIs are crucial in any Android app. A malware developer
cannot hide API access, for example \textit{SendSMS}, unless the malicious
payload is downloaded at runtime. Therefore, \textsf{PetaDroid} is resilient to
common obfuscations as long as they do not remove or hide API access calls. (2)
The other factor is fragment-randomization, which makes \textsf{PetaDroid}
models robust to code transformation and obfuscation in general. We argue that
training machine learning models on dynamic fragments enhances the resiliency of
the models against code transformation.

\begin{table}[ht!]
\centering
        \caption{Obfuscation Resiliency on DroidChameleon Dataset (\textbf{General Performance})}
\label{tab_chameleonObfuscation}

\begin{scriptsize}
\begin{tabular}{l| c c c}
\hline
\textbf{Obfuscation Techniques} & \textbf{F1 (\%)} & \textbf{P (\%)} & \textbf{R (\%)} \\
\hline
{\bf No Obfuscation                } & 99.7 & 99.8 & 99.6 \\
\hline
{\tt Class Renaming                } & 99.6 & 99.6 & 99.5 \\
{\tt Method Renaming               } & 99.7 & 99.7 & 99.7 \\
{\tt Field Renaming                } & 99.7 & 99.8 & 99.7 \\
{\tt String Encryption             } & 99.8 & 99.8 & 99.7 \\
{\tt Array Encryption              } & 99.8 & 99.8 & 99.7 \\
{\tt Call Indirection              } & 99.8 & 99.8 & 99.7 \\
{\tt Code Reordering               } & 99.8 & 99.8 & 99.7 \\
{\tt Junk Code Insertion           } & 99.8 & 99.8 & 99.7 \\
{\tt Instruction Insertion         } & 99.7 & 99.8 & 99.7 \\
{\tt Debug Information Removing    } & 99.8 & 99.8 & 99.7 \\
{\tt Disassembling and Reassembling} & 99.8 & 99.8 & 99.7 \\
\hline
\end{tabular}
\end{scriptsize}
\end{table}

\subsection{Change Over Time Resiliency}
\label{sec_petaAuto}

\begin{figure}[ht!]
\begin{center}
\subfigure[\scriptsize (2013)]{
\label{fig_detection_overtime_2013}
\includegraphics[width=.20\textwidth]{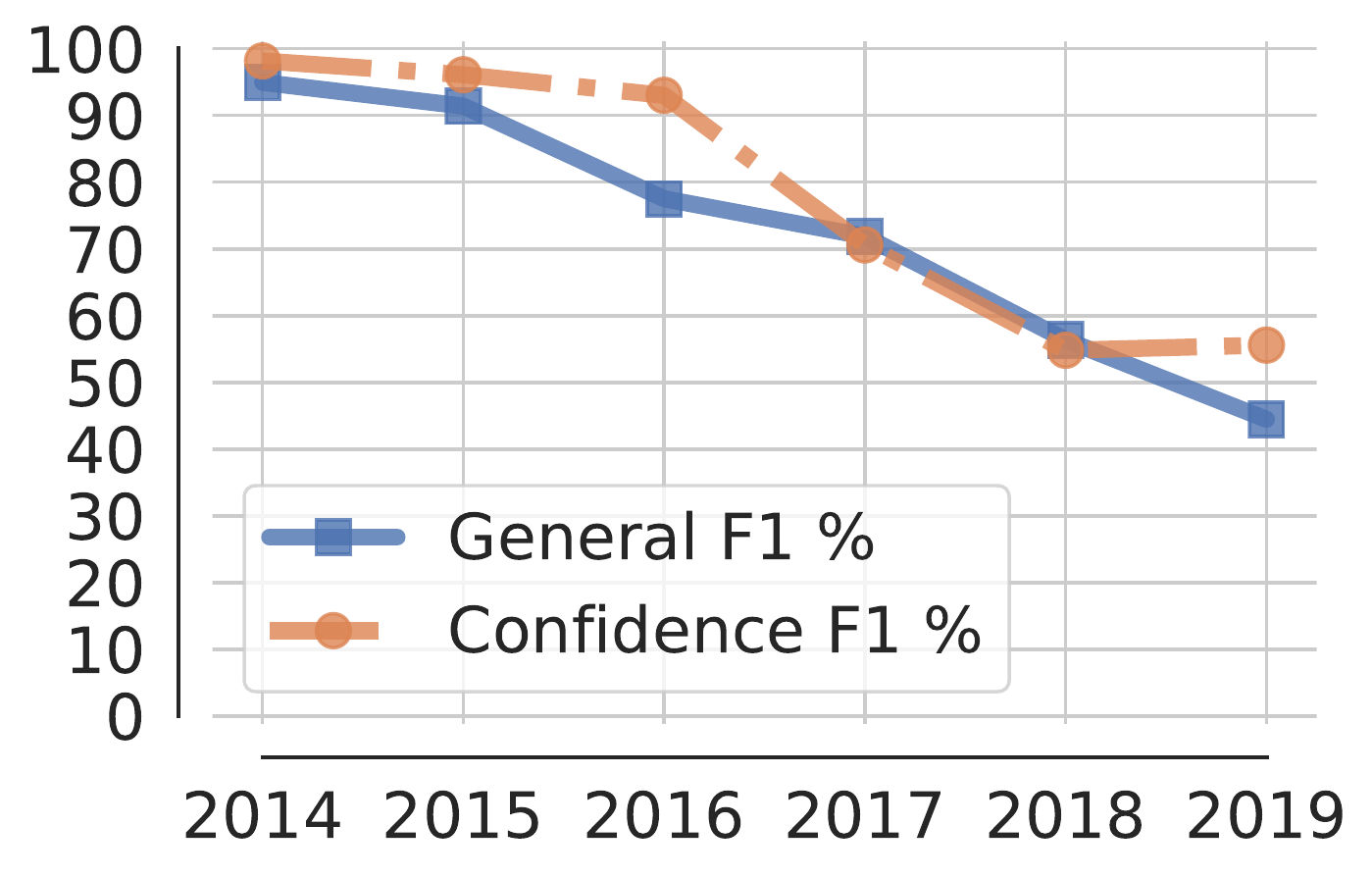}
}\\
\subfigure[\scriptsize (2014)]{
\label{fig_detection_overtime_2014}
\includegraphics[width=.20\textwidth]{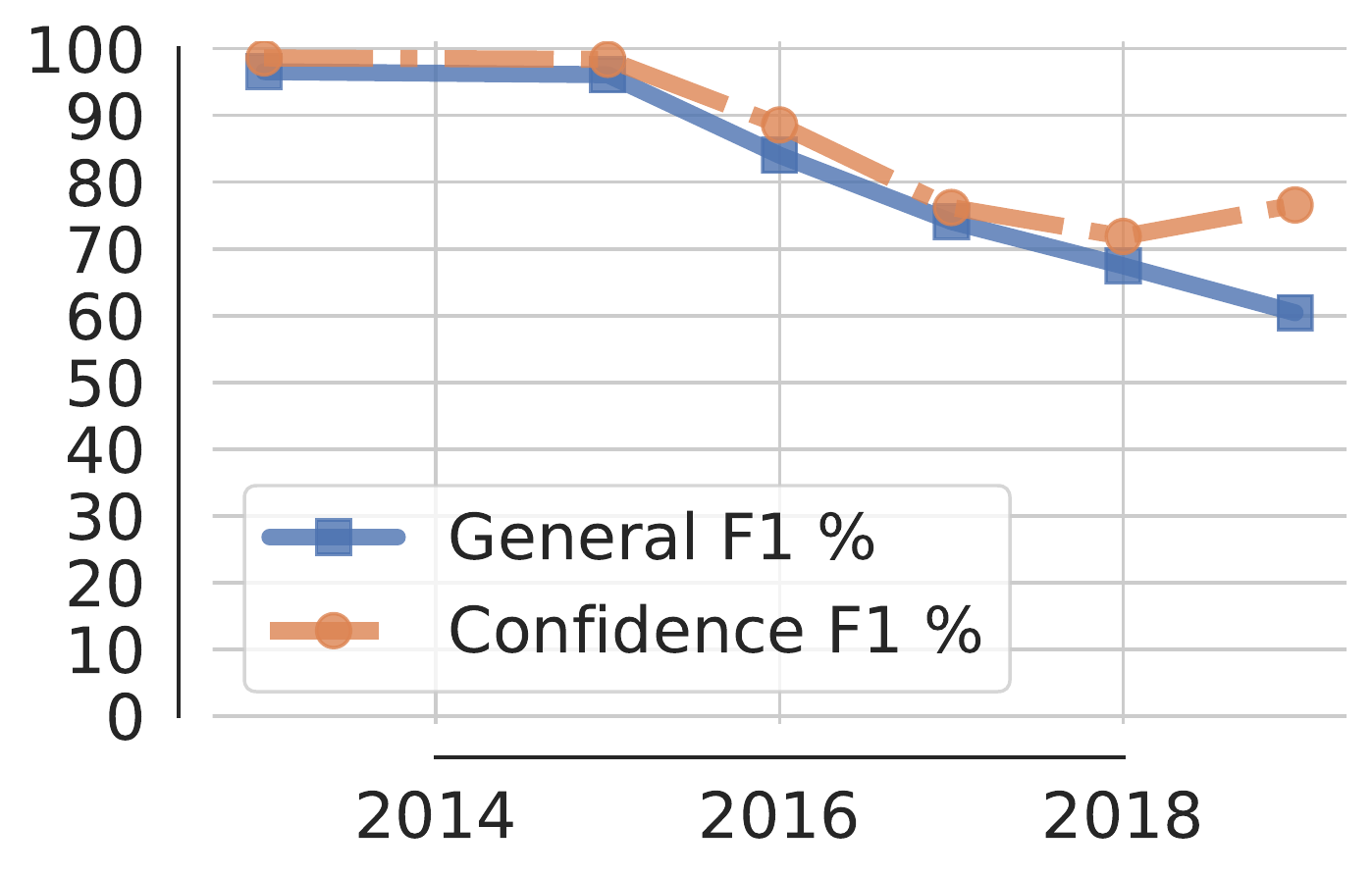}
}
\subfigure[\scriptsize (2015)]{
\label{fig_detection_overtime_2015}
\includegraphics[width=.20\textwidth]{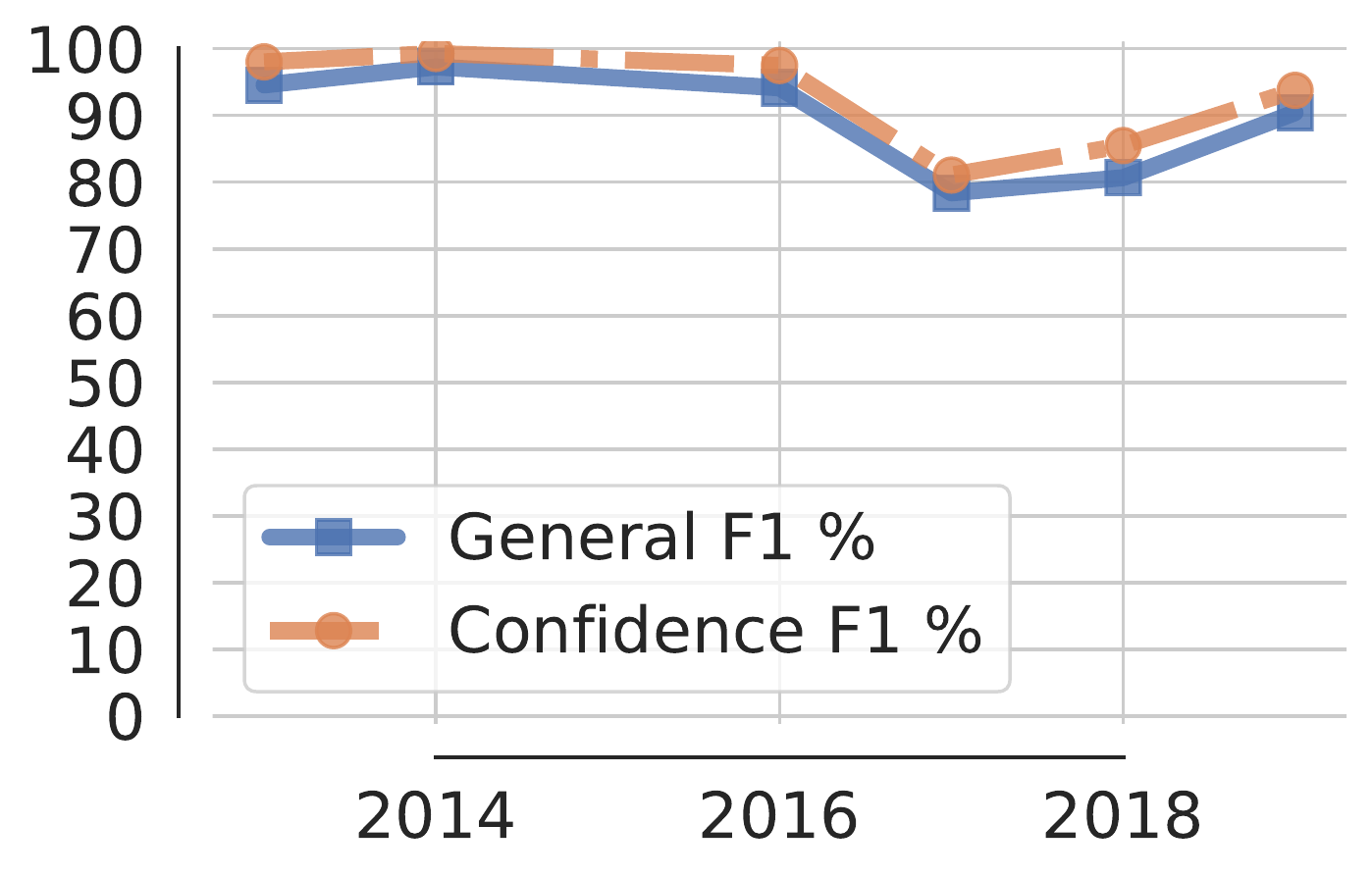}
}\\%
\subfigure[\scriptsize (2016)]{
\label{fig_detection_overtime_2016}
\includegraphics[width=.20\textwidth]{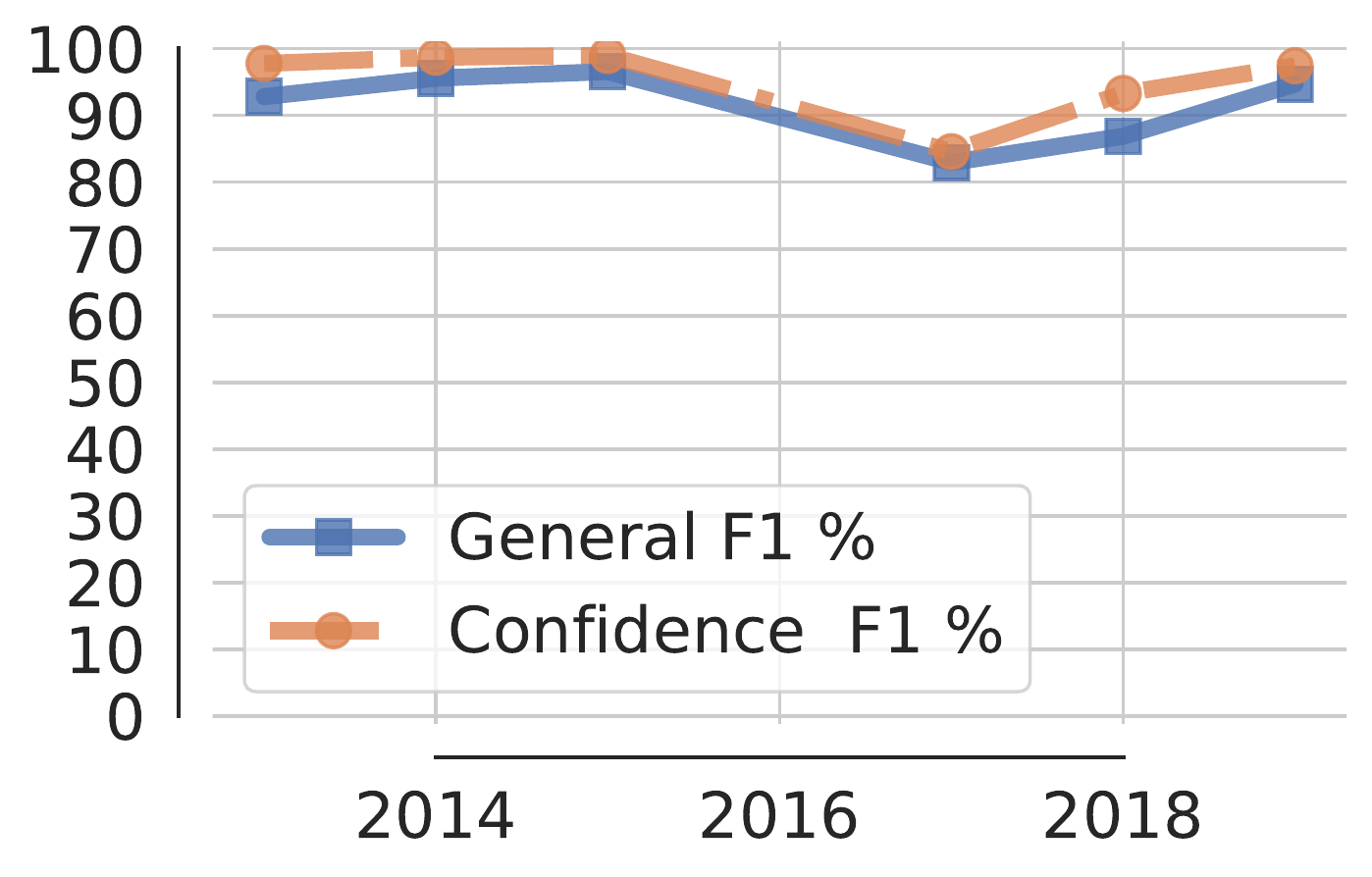}
}
\subfigure[\scriptsize (2017)]{
\label{fig_detection_overtime_2017}
\includegraphics[width=.20\textwidth]{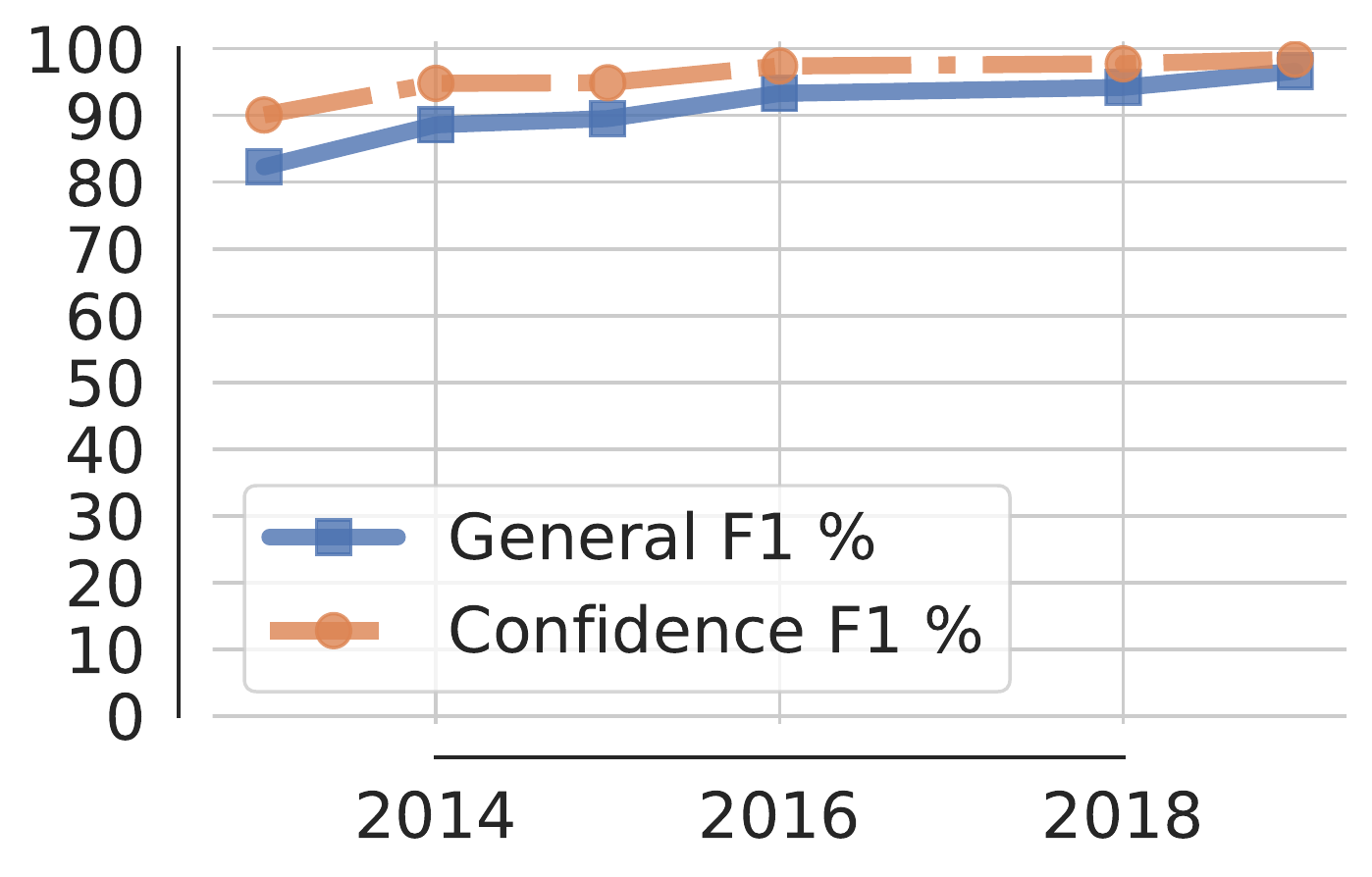}
}\\%
\subfigure[\scriptsize (2018)]{
\label{fig_detection_overtime_2018}
\includegraphics[width=.20\textwidth]{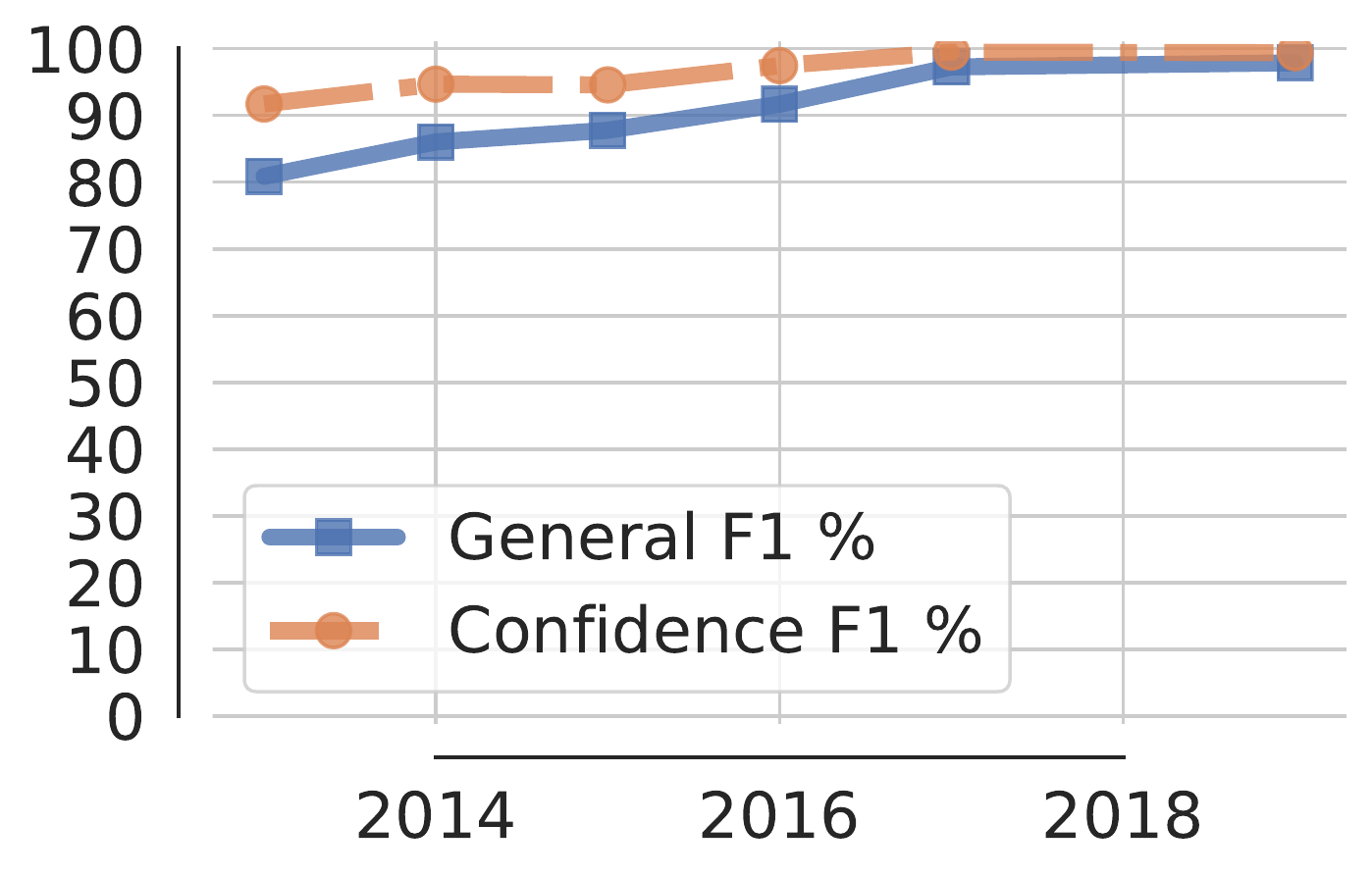}
}
\subfigure[\scriptsize (2019)]{
\label{fig_detection_overtime_2019}
\includegraphics[width=.20\textwidth]{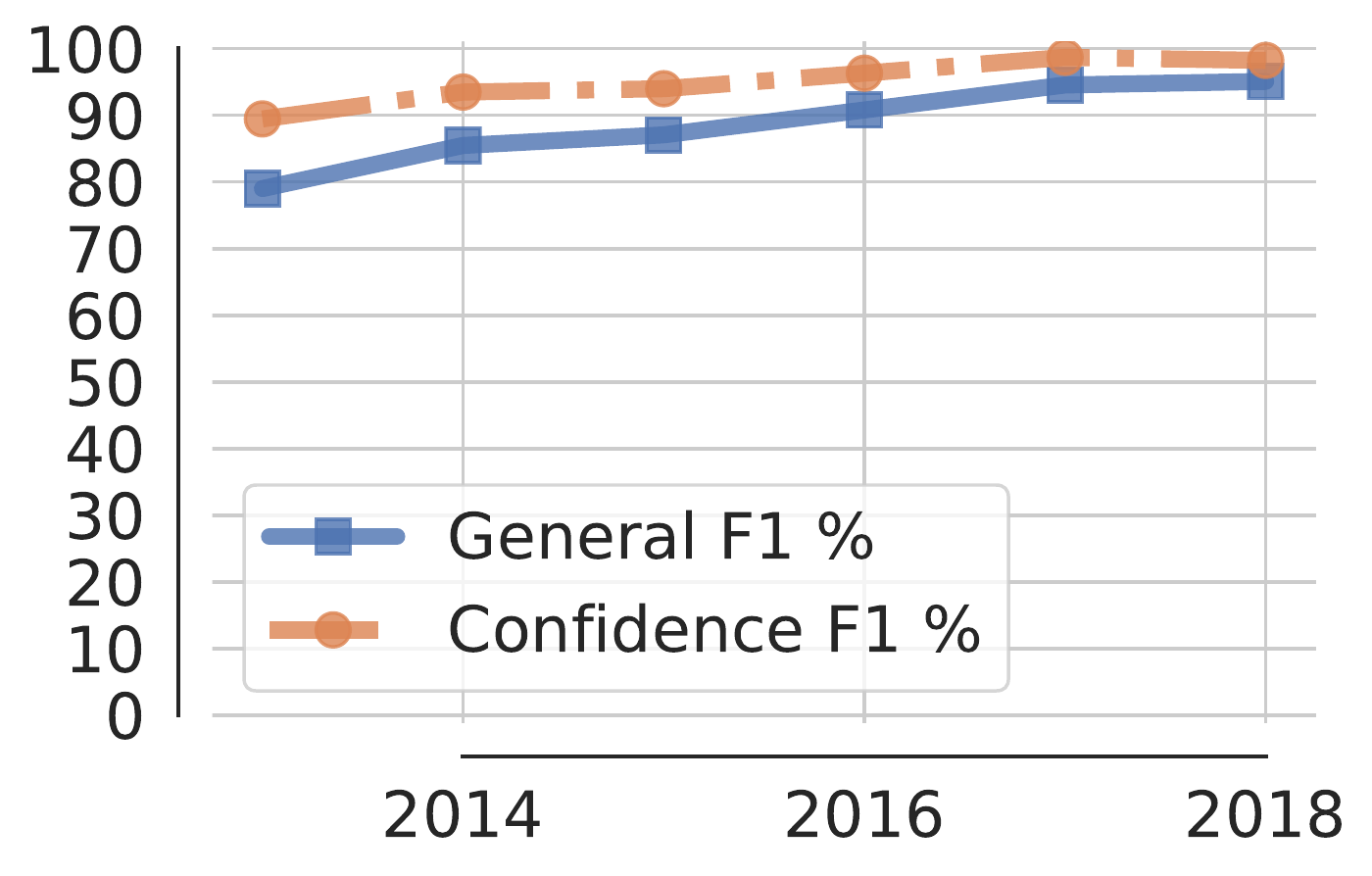}
}
\end{center}
\caption{PetaDroid Resiliency to Changes over Time}
\label{fig_detectionTimeResiliency}
\end{figure}

An important feature in modern Android malware detection is the resiliency to
change over time \cite{mariconti2017mamadroid, mariconti2019mamadroid,
karbab2018maldozer}. We study the resiliency of \textsf{PetaDroid} over the
last seven-year (2013-2019). We randomly sample from AndroZoo repository a
number of $10k$ Android apps ($5k$ malware and $5k$ benign apps) for each year
(2013-2019). As result, we have $70k=35k_{Mal}+35k_{Ben}$. We build the CNN
ensemble using year $Y_{x}$ samples and evaluate on the other years $Y_{1..N}$
samples.  Figure \ref{fig_detection_overtime_2013} shows the general and the
confidence performances of PetaDroid, for models trained on 2013 samples, in
terms of f1-score on 2014-2019 samples. As shown in Figure
\ref{fig_detection_overtime_2013}, \textsf{PetaDroid} , trained on 2013
dataset, achieved $98.17\%$, $96.10\%$, $93.01\%$, $70.60\%$, $54.82\%$,
$55.59\%$ f1-score on 2014, 2015, 2016, 2017, 2018, and 2019 datasets
respectively.  \textsf{PetaDroid} sustains a relatively good performance over
the first few years. In 2018 and 2019, the performance drops considerably. In
comparison to MaMaDroid \cite{mariconti2019mamadroid}, \textsf{PetaDroid} shows
a higher time resiliency over seven years, while MaMaDroid drops considerably
in year three ($40\%$ f1-score on year four). Figure
\ref{fig_detection_overtime_2014} shows the training is on 2014 samples, which
shows a performance enhancement over the overall evaluation period. The overall
performance tends to increase as we train on a recent year dataset as depicted
in Figure \ref{fig_detection_overtime_2015}, \ref{fig_detection_overtime_2016},
and \ref{fig_detection_overtime_2017}. In Figure
\ref{fig_detection_overtime_2018} and \ref{fig_detection_overtime_2019},
training is on samples from 2018 and 2019 respectively, \textsf{PetaDroid}
performance slightly decreases on old samples from 2013 and 2014. Our
interpretation is that old and deprecated Android APIs are not present in new
apps from 2018 and 2019, which we use for the training and this influences the
detection performance negatively. {\it We take from this experiment that
\textsf{PetaDroid} is resilient to change over time for years $t\pm2$ when we
train on year $Y_{t}$ samples.  \textsf{PetaDroid} covers about five years
$\{Y_{t-2}, Y_{t-1}, Y_{t}, Y_{t+1}, Y_{t+2}\}$  of Android app change.}

\subsection{Automatic Adaptation}
\label{sec_petaAuto}

\textsf{PetaDroid} automatic adaptation goes a step further beyond time resiliency.  \textsf{PetaDroid} employs the confidence performance to collect an extension dataset $X_{extend}$ during the deployment. \textsf{PetaDroid} automatically uses $X_{extend}$ in addition to the previous build dataset as a new build dataset $X_{build(t)}=X_{build(t-1)} \cup X_{extend}$ to build a new ensemble at every new epoch. Table \ref{tab_autonousDetection} depicts \textsf{PetaDroid} performance with and without automatic adaptation. \textsf{PetaDroid} achieves very good results compared to the previous section.
PetaDroid maintains an f1-score in the range of $83-95\%$ during all years. Without adaption, \textsf{PetaDroid} f1-score drops considerably starting from 2017. Table \ref{tab_autonousDetection} shows the performance of revisiting detection decisions on previous Android apps $X_{test}$ (benign and malware) after updating \textsf{PetaDroid} ensemble using $X_{build} \cup X_{extend}, X_{extend} \subseteq X_{test}$, where the samples in $X_{extend}$ have been removed from $X_{test}$. The update performance is significantly enhanced in the overall detection during all years.  Revisiting malware detection decisions is common practice in app markets (periodic full or partial scan the market's apps), which empowers the use case of PetaDroid automatic adaptation feature and the update metric.

\begin{table}[ht!]
\centering
\caption{Performance of PetaDroid Automatic Adaptation}
\label{tab_autonousDetection}

\begin{scriptsize}
\begin{tabular}{l|cccc}
\hline
  \textbf{Year} & \textbf{No Update(F1\%)} & \textbf{General(F1\%)} & \textbf{Confidence(F1\%)} & \textbf{Update(F1\%)}\\
\hline
\texttt{2014} & 98.2  & 97.0 & 97.9 & 99.7 \\
\texttt{2015} & 96.1  & 95.8 & 96.7 & 97.5 \\
\texttt{2016} & 93.0  & 93.3 & 94.8 & 96.4 \\
\texttt{2017} & 70.6  & 83.9 & 84.2 & 95.4 \\
\texttt{2018} & 54.8  & 87.6 & 91.6 & 93.8 \\
\texttt{2019} & 55.6  & 96.3 & 98.7 & 99.1 \\
\hline
\end{tabular}
\end{scriptsize}
\end{table}

\subsection{Efficiency}
\label{sec_petaEfficiency}
In this section, we present the average time of \textsf{PetaDroid} detection
process. The detection process includes disassembly, preprocessing, and
inference time. \textsf{PetaDroid} spends, on average, $4.0$ seconds to
fingerprint an Android app. The runtime increases for benign apps, $5.5$
seconds, because their package sizes tend to be larger compared to malicious
ones. For malware apps, \textsf{PetaDroid} spends, on average, $3.0$ seconds for
fingerprinting on the app.

\begin{figure}[ht!]
\centering
\includegraphics
[width=0.25\textwidth, trim=0cm 0cm 0cm 0cm,clip]
{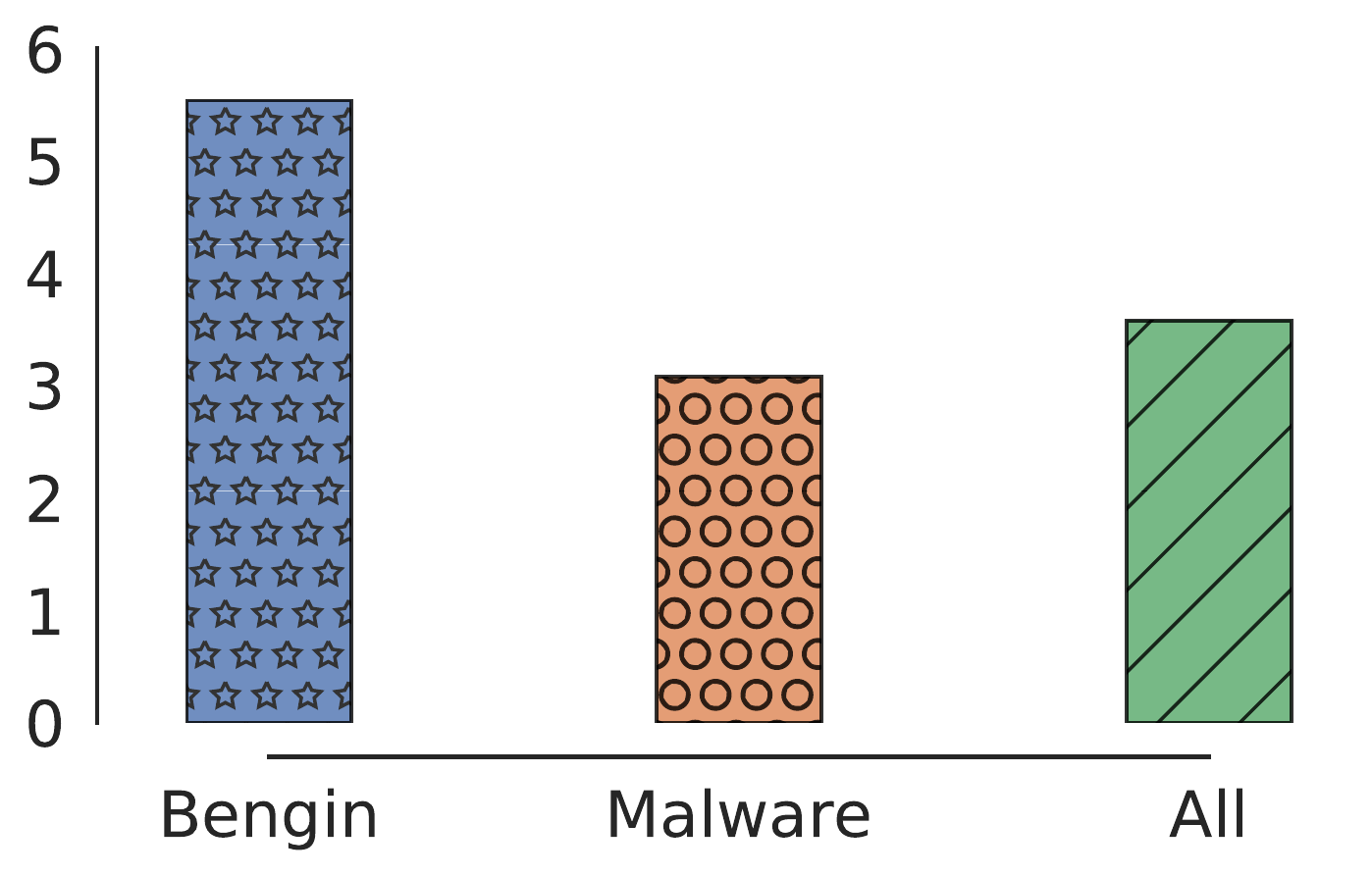}
\caption{PetaDroid Runtime Efficiency}
\label{sec_runtimeEfficiency}
\end{figure}

\section{Comparative Study}
\label{sec_compStudy}

In this section, we conduct a comparative study between \textsf{PetaDroid} and
state-of-the-art Android malware detection systems, namely: \textsf{MaMaDroid}
\cite{mariconti2017mamadroid}, \cite{mariconti2019mamadroid}, 
\textsf{DroidAPIMiner} \cite{Aafer2013DroidAPIMiner}, and {\sf MalDozer}
\cite{karbab2018maldozer}.  Our comparison is based on applying
\textsf{PetaDroid} on the same dataset (malicious and benign apps) and settings
that \textsf{MaMaDroid} used in the evaluation (provided by the authors in
\cite{mariconti2019mamadroid}). The dataset is composed of 8.5K benign and
35.5K malicious apps in addition to the Drebin \cite{arp2014drebin} dataset. The
malicious samples are tagged by time; malicious apps from 2012 (Drebin), 2013,
2014, 2015, and 2016 and benign apps are tagged as oldbenign and newbenign,
according to MaMaDroid evaluation.


\subsection{Detection Performance Comparison}
\label{sec_compDetection}

Table \ref{tab_comparativeDetection} depicts the direct comparison between
\textsf{MaMaDroid} and \textsf{PetaDroid} different dataset combinations. In
\textsf{PetaDroid}, we present the general and the confidence performance in
terms of f1-score. For \textsf{MaMaDroid} and \textsf{DroidAPIMiner}, we
present the original evaluation result \cite{mariconti2019mamadroid} in terms
of f1-score, which are equivalent to the general performance in our case.
Notice that we present only the best results of \textsf{MaMaDroid} and
\textsf{DroidAPIMiner} as reported in \cite{mariconti2019mamadroid}.

\begin{table}[ht!]
\centering
\caption{Detection Performance of MaMaDroid, PetaDroid, and DroidAPIMiner}
\label{tab_comparativeDetection}

\begin{scriptsize}
\begin{tabular}{l|ccc}
\hline
                             & Peta (F1\%)      & MaMa (F1\%) & Miner (F1\%)\\
                             &General-Confidence&             &             \\
\hline
\texttt{drebin\&oldbenign} &{\bf 98.94~-~99.40} & 96.00       & 32.00    \\
\texttt{2013\&oldbenign}   &{\bf 99.43~-~99.81} & 97.00       & 36.00    \\
\texttt{2014\&oldbenign}   &{\bf 98.94~-~99.47} & 95.00       & 62.00    \\
\texttt{2014\&newbenign}   &{\bf 99.54~-~99.83} & 99.00       & 92.00    \\
\texttt{2015\&newbenign}   &{\bf 97.98~-~98.95} & 95.00       & 77.00    \\
\texttt{2016\&newbenign}   &{\bf 97.44~-~98.60} & 92.00       & 36.00    \\
\hline
\end{tabular}
\end{scriptsize}
\end{table}

As depicted in Table \ref{tab_comparativeDetection}, \textsf{PetaDroid}
outperforms \textsf{MaMaDroid} and \textsf{DroidAPIMiner} in all datasets in
the general performance. The detection performance gap increases with the
confidence-based performance. Notice that the coverage in the confidence-based
settings is almost perfect for all the experiments in Table
\ref{tab_comparativeDetection}.

\subsection{Efficiency Comparison}
\label{sec_compEfficiency}

In Table \ref{tab_comparisonEffciency}, we report the required average time for
\textsf{MaMaDroid} and \textsf{PetaDroid} to fingerprint one Android app.
\textsf{PetaDroid} takes $03.58\pm04.21$ seconds on average for the whole
process (DEX disassembly, assembly preprocessing, CNN ensemble inference).
\textsf{MaMaDroid}, compared to \textsf{PetaDroid}, tends to be slower due to
the heavy preprocessing.  \textsf{MaMaDroid} preprocessing
\cite{mariconti2019mamadroid} is composed of the call graph extraction,
sequence extraction, and Markov change modeling, which require $25.40\pm63.00$,
$1.73\pm3.2$, $6.7\pm3.8$ seconds respectively for benign samples and
$09.20\pm14.00$, $1.67\pm3.1$, $2.5\pm3.2$ seconds respectively for malicious
samples. On average, \textsf{PetaDroid} ($3.58s$) is approximately eight times
faster than \textsf{MaMaDroid}.

\begin{table}[ht!]
\centering
\caption{MaMaDroid and PetaDroid Runtime}
\label{tab_comparisonEffciency}

\begin{scriptsize}
\begin{tabular}{lcc}
\hline 
  \textbf{} & \textbf{PetaDroid} (seconds) & \textbf{MaMaDroid} (seconds)\\
\hline
\texttt{Malware} & $02.64\pm03.94$ & $09.20\pm14.00$ + $1.67\pm3.1$ + $2.5\pm3.2$\\
\texttt{Benign } & $05.54\pm05.12$ & $25.40\pm63.00$ + $1.73\pm3.2$ + $6.7\pm3.8$\\
\texttt{Average} & $03.58\pm04.21$ &                   $\approx 23s$             \\
\hline 
\end{tabular}
\end{scriptsize}
\end{table}

\subsection{Time Resiliency Comparison}
\label{sec_compTimeResiliency}

\textsf{MaMaDroid} evaluation emphasizes the importance of time resiliency for
modern Android malware detection. Table \ref{tab_comparisonTimeResiliency}
depicts the performance with different dataset settings, such as training using
an old malware dataset and testing on a newer one. \textsf{PetaDroid}
outperforms (or obtains a very similar result in few cases) \textsf{MaMaDroid}
and \textsf{DroidAPIMiner} in all settings. Furthermore, the results show that
\textsf{PetaDroid} is more robust to time resiliency compared to
\textsf{MaMaDroid} \cite{mariconti2019mamadroid}.

\begin{table*}[h!]
\centering
\caption{Classification performance of MaMaDroid, PetaDroid, DroidAPIMiner.}
\label{tab_comparisonTimeResiliency}
\resizebox{0.999\linewidth}{!}{
\begin{tabular}{cccc|ccc|ccc|ccc|ccc}
\hline
\multicolumn{1}{c}{\bf Testing Sets} & 
\multicolumn{3}{c}{\texttt{drebin \& oldbenign}} &
\multicolumn{3}{c}{\texttt{2013   \& oldbenign}} & 
\multicolumn{3}{c}{\texttt{2014   \& oldbenign}} &
\multicolumn{3}{c}{\texttt{2015   \& oldbenign}} &
\multicolumn{3}{c}{\texttt{2016   \& oldbenign}}\\
{\bf Training Sets} &
\textbf{Miner}& \textbf{MaMa}& \textbf{Peta}& \textbf{Miner}& \textbf{MaMa}&
\textbf{Peta}& \textbf{Miner}& \textbf{MaMa}& \textbf{Peta}& \textbf{Miner}&
\textbf{MaMa}& \textbf{Peta}& \textbf{Miner}& \textbf{MaMa}& \textbf{Peta}\\
\hline
\texttt{drebin\&oldbenign} & 32.0 & 96.0 & {\bf 99.4} & 35.0 & 95.0 & {\bf 98.6} & 34.0 & 72.0& {\bf 77.5} & 30.0 & 39.0 & {\bf 44.0} & 33.0 & 42.0 & {\bf 47.0} \\
\texttt{2013\&oldbenign}   & 33.0 & 94.0 & {\bf 97.8} & 36.0 & 97.0 & {\bf 99.6} & 35.0 & 73.0& {\bf 85.4} & 31.0 & 37.0 & {\bf 59.3} & 33.0 & 28.0 & {\bf 56.6} \\
\texttt{2014\&oldbenign}   & 36.0 & 92.0 & {\bf 95.8} & 39.0 & 93.0 & {\bf 98.6} & 62.0 & 95.0& {\bf 99.4} & 33.0 & 78.0 & {\bf 91.4} & 37.0 & 75.0 & {\bf 88.9} \\
\hline
&
\multicolumn{3}{c}{\texttt{drebin \& newbenign}} &
\multicolumn{3}{c}{\texttt{2013   \& newbenign}} &
\multicolumn{3}{c}{\texttt{2014   \& newbenign}} &
\multicolumn{3}{c}{\texttt{2015   \& newbenign}} &
\multicolumn{3}{c}{\texttt{2016   \& newbenign}}\\
{\bf Training Sets} & 
\textbf{Miner}& \textbf{MaMa }& \textbf{Peta }& \textbf{Miner}& \textbf{MaMa }&
\textbf{Peta}& \textbf{Miner}& \textbf{MaMa }& \textbf{Peta }& \textbf{Miner}&
\textbf{MaMa}& \textbf{Peta }& \textbf{Miner}& \textbf{MaMa }& \textbf{Peta
  }\\
\hline
\texttt{2014\&newbenign} & 76.0 & 98.0       & {\bf 99.3} & 75.0 & 98.0 & {\bf 99.7} & 92.0 & 99.0       & {\bf 99.8} & 67.0 & 85.0 & {\bf 91.4} & 65.0 & 81.0 & {\bf 82.1} \\
\texttt{2015\&newbenign} & 68.0 & 97.0       & {\bf 97.1} & 68.0 & 97.0 & {\bf 97.8} & 69.0 & {\bf 99.0} & 98.9         & 77.0 & 95.0 & {\bf 99.0} & 65.0 & 88.0 & {\bf 95.4} \\
\texttt{2016\&newbenign} & 33.0 & {\bf 96.0} &       95.6 & 35.0 & 98.0 & {\bf 98.2} & 36.0 & {\bf 98.0} & 97.9         & 34.0 & 92.0 & {\bf 95.2} & 36.0 & 92.0 & {\bf 98.3} \\
\hline
\end{tabular}
}
\end{table*}

\subsection{PetaDroid and Maldozer Comparison}
\label{sec_petadroid_vs_maldozer}

In this section, we compare {\sf PetaDroid} with {\sf MalDozer}
\cite{karbab2018maldozer} to check the effectiveness of the proposed approach.
Specifically, we evaluate the performance of both detection systems on raw
Android datasets without any code transformation. Afterward, we evaluate the
systems on randomize code transformation. Table
\ref{tab_petadroid_vs_maldozer_effectiveness} shows the effectiveness
comparison between the detection systems. {\bf First}, {\sf PetaDroid}
outperforms {\sf MalDozer} in all the evaluation dataset without code
transformation. One major factor to this result is the usage of the machine
learning model ensemble to enhance the detection performance. {\bf Second},
this gap significantly increases when we use code transformation in the various
evaluation datasets. {\sf PetaDroid} preserves the high detection performance
due to the fragment randomization technique used in the training phase. As
depicted in Table \ref{tab_petadroid_vs_maldozer_effectiveness}, the evaluation
result shows the enhancement that the fragment randomization technique adds to
the Android malware detection overall to enhance the resiliency.

\begin{table}[ht!]
\centering
\caption{PetaDroid and MalDozer Comparison}
\label{tab_petadroid_vs_maldozer_effectiveness}
\begin{scriptsize}
\begin{tabular}{l|c|c}
\hline
          & {\bf PetaDroid (F1 \%)} & {\bf MalDozer (F1 \%)}              \\
\textbf{        }  & ~~~~~~Raw~~-~Randomization    & ~~~~~~Raw~~-~Randomization    \\
\hline                                                                        
\texttt{MalGenome} & 99.6~~-~~~~~~~99.3 & 98.1~~-~~~~~~~92.5 \\
\texttt{Drebin}    & 99.2~~-~~~~~~~99.1 & 97.4~~-~~~~~~~91.6 \\
\texttt{MalDozer}  & 98.5~~-~~~~~~~98.6 & 95.2~~-~~~~~~~89.3 \\
\texttt{AMD}       & 99.4~~-~~~~~~~99.5 & 96.1~~-~~~~~~~90.1 \\
\texttt{VShare}    & 95.8~~-~~~~~~~96.0 & 94.2~~-~~~~~~~88.1 \\
\hline
\end{tabular}
\end{scriptsize}
\end{table}

\section{Case Studies}
\label{sec_caseStudies}

In this section, we conduct market-scale experiments on AndroZoo dataset ($9.5$
million Android apps). We argue that these experiments reflect real word
deployments due to the dataset size, time distribution (2010-2019), and malware
family diversity. We report \textsf{PetaDroid}'s overall performance and
overtime performance using our automatic adaptation feature in terms of general
confidence.
\subsection{Large Scale Detection}
\label{sec_megaScale}

In this experiment, we employ $8.5$ out of $10$ million samples from AndroZoo
dataset. The used dataset is composed of $1.0$ million malicious samples and
$7.5$ millions benign sample. We filter out app samples that do not correlate
with VirusTotal, or they have less than five maliciousness flags in VirusTotal.
In our experiments, we randomly sample $k$ samples as build dataset $X_{build}$
and use the rest $8.5M - k$ as $X_{test}$. We use different $k$ sizes, $k \in
\{10k, 20k, 50k, 70k, 100k\}$, and we repeat each experiment ten times to
compute the average detection performance.

\begin{table}[H]
\centering
\begin{scriptsize}
\begin{tabular}{l| c | c}
\hline
\textbf{\#Samples} &  \textbf{General (F1 \%)} & \textbf{Confidence (F1 \%)}\\
\hline
\texttt{10k}       & 95.34                     &    97.88                   \\
\texttt{20k}       & 96.17                     &    98.01                   \\
\texttt{50k}       & 96.50                     &    98.10                   \\
\texttt{70k}       & 96.76                     &    98.11                   \\
\texttt{100k}      & 97.04                     &    98.17                   \\
\hline
\end{tabular}
\end{scriptsize}
\caption{PetaDroid Market-Scale Detection Performance}
\label{tab_megaDetection}
\end{table}

In Table \ref{tab_megaDetection}, we report the detection performance in terms of f1-score of \textsf{PetaDroid} on AndroZoo dataset. \textsf{PetaDroid} shows a high f1-score for all the experiments, $95-97\%$ f1-score. We achieved $95.34\%$ f1-score when the build set is only $10k$. We argue that fragment randomization plays an important role in achieving these detection results because it acts as a data augmenter during the training phase.

\subsection{Large Scale Automatic Adaptation}
\label{sec_megaAuto}

In this experiment, we put the automatic adaptation feature on mega-scale test using $5.5$ million samples from AndroZoo dataset (2013-2016) on 25 training epochs (every three-months). We initiate \textsf{PetaDroid} on only $25k$ build dataset collected between $2013-Jan-01$ and $2013-Jan-31$. \textsf{PetaDroid} rebuilds new CNN ensemble for each three-month samples by retraining on $X_{build(t)}=X_{build(t-1)} \cap X_{extend}$.

\begin{table}[H]

\centering
\begin{scriptsize}
\begin{tabular}{l|cc|cc}
\hline
& \multicolumn{2}{c|}{\textbf{Before Update (F1 \%)}} & \multicolumn{2}{c}{\textbf{After Update (F1 \%)}} \\
\textbf{Update Epoch} & \textbf{General} & \textbf{Confidence} & \textbf{General} & \textbf{Confidence}   \\
\hline
\texttt{2013-01-31}   &   /   &   /   &   /   &   /   \\
\texttt{2013-04-30}   & 96.02 & 98.43 & 99.01 & 99.71 \\
\texttt{2013-07-31}   & 94.52 & 96.12 & 97.95 & 99.56 \\
\texttt{2013-10-31}   & 94.42 & 97.37 & 97.03 & 99.56 \\
\texttt{2014-01-31}   & 83.45 & 92.74 & 95.45 & 99.37 \\
\texttt{2014-04-30}   & 90.48 & 96.21 & 94.15 & 99.43 \\
\texttt{2014-07-31}   & 86.98 & 95.79 & 91.53 & 99.11 \\
\texttt{2014-10-31}   & 92.32 & 98.47 & 93.11 & 99.00 \\
\texttt{2015-01-31}   & 91.57 & 97.72 & 90.91 & 99.18 \\
\texttt{2015-04-30}   & 91.31 & 98.55 & 92.72 & 99.09 \\
\texttt{2015-07-31}   & 88.16 & 97.46 & 88.90 & 98.99 \\
\texttt{2015-10-31}   & 73.82 & 87.45 & 83.44 & 97.57 \\
\texttt{2016-01-31}   & 78.59 & 90.92 & 85.11 & 96.26 \\
\texttt{2016-04-30}   & 84.78 & 95.44 & 86.38 & 98.33 \\
\texttt{2016-07-31}   & 71.39 & 88.08 & 80.27 & 93.54 \\
\texttt{2016-10-31}   & 78.31 & 85.79 & 79.68 & 90.75 \\
\hline
\end{tabular}
\end{scriptsize}
\caption{Autonomous Adaptation on a Market-Scale Dataset}
\label{tab_megaAuto}
\end{table}

In Table \ref{tab_megaAuto}, we report the general and confidence performance
before and after updating \textsf{PetaDroid} CNN ensemble on an extended build
dataset. The automatic adaption feature achieves very good results.  The
general and confidence-based performance in terms of f1-score vary between
$71.39-96.02\%$ and $85.79-98.55\%$, respectively. These performance results
increase considerably ($90.75-99.71\%$ f1-score) after revising the previous
detection decisions using an updated CNN ensemble with a new $X_{extend}$ on
each epoch.

\section{Related work}
\label{sec_relatedWork}

The Android malware analysis techniques can be classified to \textit{static
analysis}, \textit{dynamic analysis}, or \textit{hybrid analysis}. The static
analysis methods \cite{arp2014drebin}, \cite{DBLP:conf/kbse/WuLZYZ019},
\cite{DBLP:journals/jaihc/AmiraDKNK21}, \cite{karbab2016dna} use static features
that are extracted from the app, such as: requested permissions and APIs to
detect malicious app. The dynamic analysis methods \cite{canfora2016acquiring},
\cite{karbab2016dysign}, \cite{DBLP:journals/di/KarbabD19} aim to identify
behavioral signature or behavioral anomaly of the running app. These methods are
more resistant to obfuscation. The dynamic methods offer limited scalability as
they incur additional cost in terms of processing and memory. The hybrid
analysis \cite{lindorfer2014andrubis}, \cite{DBLP:conf/IEEEares/KarbabD18},
combine between both analyses to improve detection accuracy, which costs
additional computational cost. Assuming that malicious apps of the same family
share similar features, some methods \cite{kim2015structural},
\cite{karbab2016cypider}, \cite{DBLP:journals/compsec/KarbabDDM20a}, measure the
similarity between the features of two samples (similar malicious code).  The
deep learning techniques are more suitable than conventional machine learning
techniques for Android malware detection \cite{yuan2014droid}.  Research works
on deep learning for Android malware detection are recently getting more
attention \cite{karbab2018maldozer}, 
\cite{DBLP:journals/tifs/ZhangSPZNTZ20}. These deep learning models are more
venerable to common machine learning adversarial attacks as described in
\cite{DBLP:journals/tifs/0002LWW0N0020}. In contrast, {\sf PetaDroid} employs
the ensemble technique to mitigate such adversarial attacks
\cite{DBLP:conf/iclr/TramerKPGBM18} and to enhance the overall performance.  In
DroidEvolver \cite{DBLP:conf/eurosp/XuLDCX19}, the authors use online machine
learning techniques to enhance the time resiliency of the Android malware
detection system. In contrast, {\sf PetaDroid} employs batch training techniques
instead of online training, which means that in each epoch t {\sf PetaDroid}
builds new models using the extended dataset at once. We argue that batch
learning could generalize better since the training system has a complete view
of the app dataset. It is less venerable to biases that could be introduced by
the order of the apps in online training.

\textsf{PetaDroid} provides Android malware detection and family clustering
using advanced natural language processing and machine learning techniques.
\textsf{PetaDroid} is resilient to common obfuscation techniques due to code
randomization during the training. \textsf{PetaDroid} introduces a novel
automatic adaption technique inspired from
\cite{DBLP:conf/nips/Lakshminarayanan17} that leverages the result confidence to
build a new CNN ensemble on confidence detection samples. Our automatic
adaptation technique aims to overcome the issue of new Android APIs over time,
while other methods could be less resilient and might require updates with a
manually crafted dataset. The empirical comparison with state-of-the-art
solutions, {\sf MaMaDroid} \cite{mariconti2019mamadroid} and {\sf MalDozer}
\cite{karbab2018maldozer}, shows that {\sf PetaDroid} outperforms {\sf
MaMaDroid} and {\sf MalDozer} under the various evaluation settings in the
malware detection effectiveness and efficiency.


\section{Limitation}
\label{sec_PetaDroidLimitation}

Although the high obfuscation resiliency of \textsf{PetaDroid} showed in Section
\ref{sec_petaObfusResiliency}, \textsf{PetaDroid} is not immune to complex
obfuscation techniques. Also, \textsf{PetaDroid} most likely will not be
able to detect Android malware that downloads the payload during runtime.
\textsf{PetaDroid} focuses on the fingerprinting process on DEX bytecode.
Therefore, Android malware, which employs C/C++ native code, is less likely to
be detected because we do not consider native code in our fingerprinting
process. Covering native code is a possible future enhancement for
\textsf{PetaDroid}. We consider including selective dynamic analysis for low
confidence detection as future work. The latter will empower \textsf{PetaDroid}
against sophisticated obfuscation techniques.  {\sf PetaDroid} system needs
more validation on real world deployments to check the performance as proposed
in previous investigations
\cite{DBLP:conf/uss/PendleburyPJKC19,DBLP:conf/uss/JordaneySDWPNC17}. Also, we
need to check the correctness of the dataset split to prevent bias results as a
result of {\it spatial bias} and {\it temporal bias}
\cite{DBLP:conf/uss/PendleburyPJKC19}. In section \ref{sec_compTimeResiliency}
and \ref{sec_PetaDroidLimitation}, we partially addressed this issues by (1)
evaluating the system on temporal splits from AndroZoo dataset and (2)
employing collected samples dataset (VirusShare) in addition to multiple
references datasets.

\section{Conclusion}
\label{sec_conclusion}

In this paper, we presented \textsf{PetaDroid}, an Android malware detection
and family clustering framework for large scale deployments. \textsf{PetaDroid}
employs supervised machine learning, an ensemble of CNN models on top of
\textit{Inst2Vec} features, to fingerprint Android malicious apps accurately.
DBScan clustering on top of \textit{InstNGram2Vec} and deep auto-encoders
features, to cluster highly similar malicious apps into their most likely
malware family groups. 
In \textsf{PetaDroid}, we introduced fragment-based detection, in which we
randomize the macro-action of Dalvik assembly instructions while keeping the
inner order of methods' sequences. We introduced the automatic adaption
technique that leverages confidence-based decision making to build a new CNN
ensemble on confidence detection samples.
The adaptation technique automatically enhances \textsf{PetaDroid} time
resiliency.  We conducted a thorough evaluation of different reference datasets
and various settings.  We evaluate \textsf{PetaDroid} on a market scale Android
dataset, $10$ Million samples and over $100$TB of data. 
\textsf{PetaDroid} achieved high detection (98-99\% f1-score) and family
clustering (96\% cluster homogeneity) performance. Our comparative study
between \textsf{PetaDroid}, \textsf{MaMaDroid} \cite{mariconti2019mamadroid}
and {\sf MalDozer} shows that \textsf{PetaDroid}
outperforms state-of-the-art solutions on various evaluation settings.


\bibliographystyle{plain}
\bibliography{references}

\end{document}